\documentclass[11pt,aps,prd,preprintnumbers,groupedaddress,nofootinbib,amssymb,notitlepage,eqsecnum]{revtex4-1}
\usepackage{here}
\usepackage{graphicx}
\usepackage{amsmath,amsthm,amssymb}
\usepackage{bm}
\usepackage{color}

\usepackage{amsfonts}
\usepackage{dcolumn}
\usepackage{hyperref}
\allowdisplaybreaks[1]
\usepackage{stackengine}

\newcommand{\be}{\begin{equation}}  
\newcommand{\ee}{\end{equation}}
\newcommand{\ba}{\begin{eqnarray}}
\newcommand{\ea}{\end{eqnarray}}

\newcommand{\rd}{{\rm d}}

\newcommand{\bem}{\begin{bmatrix}}
\newcommand{\eem}{\end{bmatrix}}

\allowdisplaybreaks

\begin{document}

\thispagestyle{empty}
\begin{flushright}
WUCG-22-04 \\
\end{flushright}
\vspace*{1.0cm}
\begin{center}

{\Large \bf  Instability of hairy black holes in regularized}
\vskip.3cm
{\Large \bf 4-dimensional Einstein-Gauss-Bonnet gravity}  

\vspace*{1.15cm} {\large 
Shinji Tsujikawa\footnote{\tt
tsujikawa@waseda.jp} 
}\\
\vspace{.5cm}
{\em Department of Physics, Waseda University, Shinjuku, Tokyo 169-8555, Japan}\\

\vspace{1.65cm} {\bf ABSTRACT}
\end{center}

In regularized 4-dimensional Einstein-Gauss-Bonnet (EGB) gravity derived from 
a Kaluza-Klein reduction of higher-dimensional EGB theory, 
we study the existence and stability of black hole (BH) solutions on a static 
and spherically symmetric background. 
We show that asymptotically-flat hairy BH solutions realized for 
a spatially-flat maximally symmetric internal space are unstable 
against linear perturbations for any rescaled GB coupling constant. 
This instability is present for the angular propagation of even-parity perturbations 
both in the vicinity of an event horizon and at spatial infinity. 
There is also a strong coupling problem associated with the 
kinetic term of even-parity perturbations vanishing everywhere.

\vfill \setcounter{page}{0} \setcounter{footnote}{0}

\vspace{1cm}
\newpage

\vspace{1cm}

\section{Introduction}

There have been numerous attempts for constructing gravitational 
theories beyond General Relativity 
(GR) \cite{Copeland:2006wr,DeFelice:2010aj,Clifton:2011jh,Joyce:2014kja,Heisenberg:2018vsk}. 
This is not only motivated by the construction of an ultraviolet compete theory 
of gravity but also by the firm observational evidence of dark matter 
and dark energy. Moreover, the dawn of gravitational-wave 
astronomy \cite{LIGOScientific:2016aoc} started 
to offer a new opportunity for probing the physics of strong gravity regimes 
in the vicinity of black holes (BHs) and neutron stars (NSs) \cite{Berti:2015itd,Barack:2018yly}.
It is important to scrutinize the possible deviation from GR 
on the cosmological and strong gravitational backgrounds. 

GR is the 4-dimensional spacetime theory that leads to conserved and 
symmetric field equations of motion with derivatives up to second order 
in the metric tensor. One way of modifying the gravitational equations in GR 
is to add new degrees of freedom such as scalar/vector fields coupled to gravity. 
For example, Horndeski theories \cite{Horndeski,Def11,KYY,Charmousis:2011bf} 
are the most general scalar-tensor theories whose equations of motion 
contain derivatives of the scalar field and metric tensor up to second order. 
In such second-order gravitational theories, one can avoid an Ostrogradski
instability \cite{Ostrogradsky:1850fid} arising from the Hamiltonian unbounded from below. 

The other way of constructing alternative theories of gravity is 
to consider spacetime dimensions higher than 4. 
In this vein, Lovelock \cite{Lovelock:1971yv} derived the gravitational 
field equations of motion with 2-rank symmetric tensors satisfying 
the conserved and second-order properties.
In 4 dimensions, GR is the unique theory of gravity 
being compatible with such properties \cite{Lovelock:1972vz}. 
In spacetime dimensions higher than 4, there is a new term 
known as the Gauss-Bonnet (GB) term defined by 
\be
{\cal R}_{\rm GB}^2={\cal R}^2-4{\cal R}_{AB}{\cal R}^{AB}
+{\cal R}_{ABCD}{\cal R}^{ABCD}\,,
\ee
where ${\cal R}$, ${\cal R}_{AB}$, and ${\cal R}_{ABCD}$ are
the Ricci scalar, Ricci tensor, and Riemann tensors in 
$D>4$ dimensions, respectively \cite{Padmanabhan:2013xyr,Fernandes:2022zrq}. 
In 4 spacetime dimensions, the field equations of motion 
following from the Lagrangian $L=\sqrt{-g}\,\hat{\alpha} {\cal R}_{\rm GB}^2$ 
(where $g$ is a determinant of the metric tensor and
$\hat{\alpha}$ is a GB coupling) identically vanish by 
reflecting the fact that the GB curvature invariant 
reduces to a topological surface term.

If we rescale the GB coupling constant as $\hat{\alpha} \to \alpha/(D-4)$ 
and take the limit $D \to 4$, it is possible to extract non-trivial contributions 
of the higher-dimensional GB term in 4 dimensions \cite{Glavan:2019inb}. 
This theory is known as 4D-Einstein-Gauss-Bonnet (4DEGB) gravity, 
which has received much attention over the past 
few years (see Ref.~\cite{Fernandes:2022zrq} for a review). 
In the original version proposed by Glavan and Lin \cite{Glavan:2019inb}, 
there are some ill-defined terms in the field equations that lead to the divergence of second-order perturbation equations about a Minkowski 
background \cite{Arrechea:2020evj,Arrechea:2020gjw}. 
Furthermore, the $D \to 4$ limit gives rise to unphysical divergences 
in the on-shell action \cite{Mahapatra:2020rds}. 
In Refs.~\cite{Gurses:2020ofy,Gurses:2020rxb}, 
it was also argued that a simple rescaling of 
the GB coupling originally suggested by Glavan and Lin does not 
yield covariant equations of a massless graviton in 
4 dimensions.

However, there are several possibilities for constructing regularized 4DEGB 
theories that can evade the aforementioned problems of original 
4DEGB gravity.
One of such improved theories arises from a conformal regularization 
introducing a counter-term to eliminate 
divergent parts of the theory \cite{Fernandes:2020nbq,Hennigar:2020lsl}.
The other is a Kaluza-Klein reduction of the higher-dimensional 
EGB theory with a scalar field $\phi$ characterizing the size of a 
maximally symmetric internal space \cite{Lu:2020iav,Kobayashi:2020wqy}. 
In the former and latter approaches, the 4-dimensional actions 
obtained from the rescaling $\hat{\alpha} \to \alpha/(D-4)$ 
belong to subclasses of shift-symmetric and 
non-shift-symmetric Horndeski theories, 
respectively. For a spatially-flat internal space with the 
Kaluza-Klein reduction, the 4-dimensional effective action
coincides with the one derived from the conformal regularization. 
There is yet another regularization scheme in which the 
temporal diffeomorphism invariance is explicitly 
broken \cite{Aoki:2020lig}.

The regularized 4DEGB theories have been applied to the search for 
static and spherically symmetric BH solutions. 
In the original 4DEGB gravity, there exists an exact BH 
solution \cite{Glavan:2019inb} analogous to the one 
obtained in the framework of higher-dimensional EGB 
theories \cite{Boulware:1985wk,Wheeler:1985qd} 
(see also Refs.~\cite{Konoplya:2020bxa,Guo:2020zmf,Fernandes:2020rpa,Konoplya:2020qqh,Hegde:2020xlv,Ghosh:2020vpc,Ghosh:2020syx,HosseiniMansoori:2020yfj,Wei:2020poh,Mishra:2020gce,Singh:2020xju,EslamPanah:2020hoj,Yang:2020czk,Zeng:2020dco,Dadhich:2020ukj,Shaymatov:2020yte,Aragon:2020qdc,Atamurotov:2021imh,Sengupta:2021mpf,Jaryal:2022rzd,Bravo-Gaete:2022mnr} 
for related works).
The same kind of solutions is also present in two regularized versions of 
4DEGB theories mentioned above \cite{Lu:2020iav,Fernandes:2021ysi}. 
In the conformal regularization, this is the unique asymptotically-flat 
BH solution allowed in the theory. 

To study the linear stability of static and spherically symmetric BHs 
in 4 dimensions, we need to classify perturbations into the odd- and 
even-parity sectors depending on the types of parities under the rotation 
in two-dimensional sphere $(\theta, \varphi)$ \cite{Regge:1957td,Zerilli:1970se}.
In Horndeski theories, the BH stability  
against odd- and even-parity perturbations for a time-independent 
scalar field was originally addressed by Kobayashi {\it et al.} \cite{Kobayashi:2012kh,Kobayashi:2014wsa} 
(see also Refs.~\cite{DeFelice:2011ka,Motohashi:2011pw,Kase:2014baa,Tattersall:2018nve,Kase:2020qvz,Khoury:2020aya}).
Except for the propagation of even-parity modes along the 
angular direction, Kobayashi {\it et al.} derived conditions for the absence 
of ghost/Laplacian instabilities in high frequency limits.
In 4DEGB gravity, the stability of BHs has been discussed
in Refs.~\cite{Konoplya:2020juj,Cuyubamba:2020moe,Langlois:2022eta,Langlois:2022ulw} 
by paying particular attention to the behavior of perturbations 
along the radial direction.  

The recent general analysis of BH and NS perturbations in Horndeski 
theories \cite{Kase:2021mix} 
shows that the angular propagation of even-parity modes 
with large multipoles $\ell \gg 1$ 
plays an important role to exclude some hairy BH solutions by the 
instability in the vicinity of an event horizon \cite{Minamitsuji:2022mlv}.
This instability can arise for BHs with static scalar hair whose kinetic 
term $X$ is a non-vanishing constant $X_s$ on the horizon \cite{Minamitsuji:2022vbi}. 
For example, the derivative coupling with an Einstein tensor gives rise to   
exact hairy BH solutions \cite{Rinaldi:2012vy,Anabalon:2013oea,Minamitsuji:2013ura} 
with $X_s \neq 0$, but they are prone to either ghost or Laplacian instabilities. 
To recognize such properties, the propagation speeds of even-parity perturbations
for large angular momentum modes are crucial.

In this paper, we study the existence and stability of BHs
against odd- and even-parity perturbations in regularized 4DEGB gravity 
derived from the Kaluza-Klein reduction of higher-dimensional EGB theory.
Our analysis also accommodates hairy BHs present in 4DEGB theory  
with the conformal regularization. We will show that asymptotically-flat hairy 
BH solutions present in such regularized 4DEGB theories are excluded by 
the linear instability and strong coupling problems.

\section{Black hole solutions in regularized 4DEGB theories}
\label{Sec:back}

We first briefly revisit the regularized Kaluza-Klein reduction of $D$-dimensional 
EGB gravity on a $(D-4)$-dimensional maximally symmetric space 
with constant curvature $\lambda$.
In this setup, the $D$-dimensional metric is expressed in the form  
\be
{\rm d}s_D^2={\rm d} s_4^2+{\rm e}^{-2\phi}
{\rm d}\sigma_{D-4}^2\,,
\label{metric}
\ee
where ${\rm d}s_4^2$ is the line element of 4-dimensional spacetime, 
and ${\rm d}\sigma_{D-4}^2$ is the line element of an internal space 
whose Riemann tensor is given by 
${\cal R}_{abcd}=\lambda (g_{ac} g_{bd}-g_{ad} g_{bc})$. 
The scalar field $\phi$, which characterizes the size of internal space, 
depends only on the 4-dimensional 
coordinates $x^{\mu}=(x^0,x^1,x^2,x^3)$.

We consider the action of $D$-dimensional EGB theory 
given by 
\be
{\cal S}=\frac{1}{16 \pi G_D} \int {\rm d}^D x \sqrt{-g} \left( 
{\cal R}+\hat{\alpha}{\cal R}_{\rm GB}^2 \right)\,,
\label{SD}
\ee
where $G_D$ is the gravitational constant in $D$-dimensions.
Substituting the metric ansatz (\ref{metric}) into Eq.~(\ref{SD}) 
and performing the integration by parts, we obtain the following 
reduced action in 4 dimensions \cite{Lu:2020iav,Kobayashi:2020wqy} 
(see also Refs.~\cite{VanAcoleyen:2011mj,Charmousis:2012dw}):
\ba
{\cal S}
&=& \frac{1}{16 \pi G} \int {\rm d}^4 x \sqrt{-g}\,{\rm e}^{-(D-4)\phi}
\biggl[ R-(D-4)(D-5) \left( 2X-\lambda {\rm e}^{2\phi} \right)
+\hat{\alpha} R_{\rm GB}^2 \nonumber \\
& &
-2\hat{\alpha}(D-4)(D-5)
\left( 2G^{\mu \nu} \nabla_{\mu} \phi \nabla_{\nu} \phi
-\lambda R {\rm e}^{2\phi} \right) \nonumber \\
& &-4\hat{\alpha}(D-4)(D-5)(D-6) \left\{ X \square \phi+(D-5)X^2 \right\}
\nonumber \\
& &
-\hat{\alpha}(D-4)(D-5)(D-6)(D-7) \left( 4\lambda X {\rm e}^{2\phi}
-\lambda^2 {\rm e}^{4\phi} \right) \biggr]\,,
\label{SD2}
\ea
where $R$, $G^{\mu \nu}$, and $R_{\rm GB}^2$ are the Ricci scalar, 
Einstein tensor, and GB term in 4 dimensions, respectively, and 
$X=-(1/2)\nabla^{\mu}\phi \nabla_{\mu}\phi$ is the scalar kinetic 
term, and $\nabla_{\mu}$ is a covariant derivative operator.
We performed a volume integral with respect to the internal space 
and absorbed the integration constant into $G_D$ to define 
the 4-dimensional gravitational constant $G$. 
Since the GB combination is topological in 4 dimensions, we can 
add a counter term $-(16 \pi G)^{-1} \int  {\rm d}^4 x \sqrt{-g}\,
\hat{\alpha}R_{\rm GB}^2$ to the action (\ref{SD2}).
To extract contributions of the higher-dimensional GB term, we rescale 
the coupling constant to be $\hat{\alpha}=\alpha/(D-4)$ as in the 
original prescription of Glavan and Lin. 
Taking the $D \to 4$ limit at the end, we obtain the action of regularized 
4DEGB theory with the Kaluza-Klein reduction in the form 
\be
{\cal S}=\int {\rm d}^4 x \sqrt{-g} \left[ R 
+\alpha \left\{ 4 G^{\mu \nu} \nabla_{\mu} \phi  \nabla_{\nu} \phi 
-\phi R_{\rm GB}^2-8 X \square \phi+8 X^2
-2\lambda {\rm e}^{2\phi} \left( R-12 X+3\lambda 
{\rm e}^{2\phi} \right) 
\right\} \right]\,.
\label{actionre}
\ee
Here and in the following, we use the unit $(16 \pi G)^{-1}=1$.
For the spatially-flat internal space ($\lambda=0$), the action (\ref{actionre}) recovers 
the one derived by a counter term regularization with 
a conformal rescaling of the metric \cite{Fernandes:2020nbq,Hennigar:2020lsl}.

The regularized 4DEGB gravity (\ref{actionre}) belongs to a subclass of 
Horndeski theories given by the action \cite{Horndeski,Def11,KYY,Charmousis:2011bf} 
\ba
{\cal S}
&=&\int {\rm d}^4 x \sqrt{-g}\,
\biggl[ G_2-G_{3}\square\phi 
+G_{4} R +G_{4,X} \left\{ (\square \phi)^{2}
-(\nabla_{\mu}\nabla_{\nu} \phi)
(\nabla^{\mu}\nabla^{\nu} \phi) \right\}
+G_{5} G^{\mu \nu} \nabla_{\mu}\nabla_{\nu} \phi \nonumber \\
&&
-\frac{1}{6}G_{5,X}
\left\{ (\square \phi )^{3}-3(\square \phi)\,
(\nabla_{\mu}\nabla_{\nu} \phi)
(\nabla^{\mu}\nabla^{\nu} \phi)
+2(\nabla^{\mu}\nabla_{\alpha} \phi)
(\nabla^{\alpha}\nabla_{\beta} \phi)
(\nabla^{\beta}\nabla_{\mu} \phi) \right\} \biggr]\,,
\label{action}
\ea
where $G_{j,X} \equiv \partial G_j/\partial X$ (with $j=4,5$), 
and 
\ba
& &
G_2=\alpha \left( 8X^2+24 \lambda X {\rm e}^{2\phi}
-6\lambda^2 {\rm e}^{4\phi} \right)\,,\qquad 
G_3=8\alpha X\,,\nonumber \\
& & 
G_4=1+4\alpha X-2\alpha \lambda {\rm e}^{2\phi}\,,\qquad 
G_5=4\alpha \ln |X| \,.
\label{G2345}
\ea
Here, we used the fact that the term $-4\alpha \phi$ in $G_5$ is 
equivalent to the derivative coupling $4 \alpha X$ 
in $G_4$ \cite{KYY}. 
For $\lambda=0$, the functions $G_{2,3,4,5}$ depend on $X$ alone. 
In this case, the theory belongs to a subclass of shift-symmetric 
Horndeski theories whose equations of motion are invariant 
under the shift $\phi \to \phi+c$. 
For $\lambda \neq 0$, the shift symmetry is explicitly broken.

To study BH solutions in regularized 4DEGB theory given by 
the action (\ref{action}), we consider the static and spherically 
symmetric background with the line element
\be 
\rd s^2=-f(r) \rd t^{2} +h^{-1}(r) \rd r^{2}
+ r^{2} \left( \rd \theta^{2}+\sin^{2}\theta\,\rd\varphi^{2} 
\right)\,,
\label{background}
\ee
where $f$ and $h$ are functions of the radial coordinate~$r$. 
The scalar field is assumed to be a function of $r$. 
Computing (\ref{action}) on the background (\ref{background})
and varying the reduced action with respect to $f$, $h$, and $\phi$,  
respectively, we obtain the following equations 
\ba
& &
f'= -f\frac{r^2 (h-1)+\alpha [h^2-2h (1-2j-2r \phi' j)
+1-4j+3j^2]}{hr[r^2-2\alpha(h-1+j+r\phi'j)]}\,,
\label{feq}\\
& &
\frac{h'}{h}-\frac{f'}{f}=-\frac{4\alpha rj (\phi'^2-\phi'')}
{r^2-2\alpha(h-1+j+r \phi' j)}\,,
\label{heq} \\
& &
J'=\frac{2\alpha \lambda {\rm e}^{2\phi}
[r(j'+\phi' j)+j]}{1+r\phi'} \sqrt{\frac{f}{h}}\,,
\label{Jeq}
\ea
where a prime represents the derivative 
with respect to $r$, and 
\be
j \equiv 1-h (1+r \phi')^2+\lambda r^2 {\rm e}^{2\phi}\,,\qquad
J \equiv \sqrt{\frac{h}{f}}\alpha \left( f'+2\phi' f \right) j\,.
\ee
There is no standard kinetic term $X$ in the couplings functions 
(\ref{G2345}), so both the left and right hand-sides of 
Eq.~(\ref{Jeq}) vanish in the limit $\alpha \to 0$.

We search for hairy asymptotically-flat BH solutions where the metric 
components and scalar field have the following dependence 
at large distances
\be
f=f_0+\frac{f_1}{r}+\frac{f_2}{r^2}\cdots\,,\qquad 
h=1+\frac{h_1}{r}+\frac{h_2}{r^2}\cdots\,,\qquad
\phi=\phi_0+\frac{\phi_1}{r}+\frac{\phi_2}{r^2}\cdots\,, 
\label{fhexpan}
\ee
where $f_j$, $h_j$, and $\phi_j$ ($j=0,1,2,\cdots$) are constants. 
With this boundary condition of $\phi$, the size of internal space 
approaches a constant value $\phi_0$ as $r \to \infty$.
Then, at large distances, the leading-order contribution to $j$ 
for $\lambda \neq 0$ is given by $\lambda r^2 {\rm e}^{2\phi_0}$, 
so the right hand-side of Eq.~(\ref{Jeq}) has the dependence 
$6 \alpha \lambda^2 {\rm e}^{4\phi_0} f_0^{1/2} r^2$. 
Meanwhile, the term $J$ behaves as 
$J=-\alpha (2f_0 \phi_1+f_1)f_0^{-1/2} \lambda {\rm e}^{2\phi_0}
+{\cal O}(1/r)$. Then, the consistency of 
Eq.~(\ref{Jeq}) at spatial infinity leads to 
\be
\lambda=0\,.
\ee
In other words the asymptotic flatness requires 
the spatially-flat internal space, so we consider 
BH solutions in this case.

For $\lambda=0$ the right hand-side of Eq.~(\ref{Jeq}) vanishes, 
and hence $J=C={\rm constant}$. 
On using the expansions (\ref{fhexpan}), we have 
$J=\sqrt{h/f}\,\alpha( f'+2\phi' f) [1-h (1+r \phi')^2] \to 0$ 
at spatial infinity. To satisfy this boundary condition, 
we require that $C=0$ and hence 
\be
J=\sqrt{\frac{h}{f}}\alpha \left( f'+2\phi' f \right)j=0\,.
\label{Jeq2}
\ee
Since $f$, $h$, $f'$, and $\phi'$ should be finite 
throughout the horizon exterior, it follows that 
\be
j=1-h (1+r \phi')^2=0\,.
\label{j=0}
\ee
Among the two branches of Eq.~(\ref{j=0}), the solution 
that behaves as $\phi' \propto 1/r^2$ at spatial infinity is 
\be
\phi'=\frac{1}{r} \left( \frac{1}{\sqrt{h}}-1 \right)\,,
\label{phid}
\ee
whose existence requires that $h>0$.
The other branch, $\phi'=(1/r)(-1/\sqrt{h}-1)$, has the asymptotic 
behavior $\phi' \simeq -2/r$ as $r \to \infty$, so the scalar field 
exhibits a logarithmic divergence $\phi \to -2 \ln r$. 
Hence we take the branch (\ref{phid}) as a physically 
acceptable solution. For $\alpha \neq 0$, we do not have
the no-hair branch $\phi'=0$.
This is an outcome of the absence of a standard kinetic term $X$ 
in the coupling functions (\ref{G2345}).

{}From Eq.~(\ref{phid}), the field kinetic term is given by  
\be
X=-\frac{(1-\sqrt{h})^2}{2r^2}\,. 
\label{Xf}
\ee
On the horizon characterized by the radius $r_s$, 
$X$ has a non-vanishing value $X_s=-1/(2r_s^2)$, 
with the divergent field derivative $\phi' \propto 1/(r\sqrt{h})$.
The metric components around $r=r_s$ can be 
expanded as $f=\hat{f}_1(r-r_s)+\hat{f}_2 (r-r_s)^2+\cdots$ and 
$h=\hat{h}_1(r-r_s)+\hat{h}_2 (r-r_s)^2+\cdots$. 
Then, the term $\phi' f$ in $J$ vanishes at $r=r_s$, so $j=0$ 
is the solution to $J=0$ even on the horizon.
Substituting $j=0$ into Eq.~(\ref{heq}), we 
obtain the solution $h=C_1 f$, where $C_1$ is a constant. 
With a suitable time reparametrization of $f$, 
we can choose $C_1=1$. Then, the differential Eq.~(\ref{feq}) 
reduces to 
\be
f'=-\frac{(f-1)[r^2+\alpha (f-1)]}
{r [r^2 -2\alpha (f-1)]}=h'\,.
\label{dfh}
\ee
The integrated solution to this equation, which is consistent 
with the boundary conditions (\ref{fhexpan}) at spatial infinity,  
is given by 
\be
f=h=1+\frac{r^2}{2\alpha} \left[ 1-\sqrt{1+\frac{8\alpha M}{r^3}}
\right]\,,
\label{fhal}
\ee
where $M$ is an integration constant. 
While we are considering regularized 4DEGB gravity given by the effective 
action (\ref{actionre}), the BH solution with same metric components 
as (\ref{fhal}) was also found in the original 4DEGB theory \cite{Glavan:2019inb}.
At large distances metric components behave as 
$f=h=1-2M/r+\cdots$, so the field derivative 
(\ref{phid}) decreases as $\phi' \simeq M/r^2$.
For a small coupling $\alpha$, expanding Eq.~(\ref{fhal}) 
with respect to $\alpha$ leads to 
$f=h=1-2M/r+4\alpha M^2/r^4+\cdots$. 
In the limit $\alpha \to 0$, Eq.~(\ref{fhal}) reduces to 
the Schwarzschild metric.

{}From Eq.~(\ref{fhal}), there are two horizons located at 
\be
r_{\pm}=M \pm \sqrt{M^2-\alpha}\,.
\ee
The outer horizon $r_{+}$ exists for the couplings
\be
-8 M^2<\alpha<M^2\,.
\label{alrange}
\ee
In the coupling range $0<\alpha<M^2$, we have $f>0$ 
for $0 \le r<r_{-}$ and $r_{+}<r$, while $f<0$ for 
$r_{-}<r<r_{+}$.  
In the range $-8 M^2<\alpha<0$, metric components 
(\ref{fhal}) become imaginary for $r<r_* \equiv (-8M \alpha)^{1/3}$, 
so only the region $r \geq r_*$ is physically meaningful. 
Since $r_{+} \geq r_*$, 
$r_{*}$ is located inside the outer horizon $r_{+}$. 
For $\alpha=M^2$ there is only the single horizon at 
$r_{\pm}=M$, in which case $f \ge 0$ at any radius $r$. 

For $\lambda=0$, we thus showed the uniqueness of 
BH solution (\ref{fhal}) with the field derivative (\ref{phid}).
The same conclusion was also reached in Ref.~\cite{Fernandes:2021ysi}.
We have also seen that, for $\lambda \neq 0$, imposing the asymptotic 
flatness of solutions is not consistent with the scalar-field Eq.~(\ref{Jeq}).
 
\section{Instability of hairy black holes}
\label{Sec:per}

We now address the linear stability of asymptotically-flat hairy BHs 
present for $\lambda=0$. We study the behavior of odd- and even-parity 
perturbations on the background solution (\ref{fhal}) with (\ref{phid}).
The quantities relevant to the stability against odd-parity perturbations 
are given by \cite{Kobayashi:2012kh}
\ba
{\cal G} &\equiv& 2G_4+2 h\phi'^2G_{4,X}
-h\phi'^2 \left( G_{5,\phi}+{\frac {f' h\phi' G_{5,X}}{2f}} \right)\,,
\label{cGdef}\\
{\cal F} &\equiv& 2G_4+h\phi'^2G_{5,\phi}-h\phi'^2  \left( \frac12 h' \phi'+h \phi'' \right) 
G_{5,X}\,,
\label{cFdef}\\
{\cal H} &\equiv& 2G_4+2 h\phi'^2G_{4,X}-h\phi'^2G_{5,\phi}
-\frac{h^2 \phi'^3 G_{5,X}}{r}\,,
\label{cHdef}
\ea
where the no-ghost condition corresponds to ${\cal G}>0$. 
The squared propagation speeds of odd-parity gravitational perturbations 
along the radial and angular directions are given, respectively, by 
$c_{r,{\rm odd}}^2={\cal G}/{\cal F}$ and 
$c_{\Omega,{\rm odd}}^2={\cal G}/{\cal H}$. 
Under the no-ghost condition ${\cal G}>0$, the Laplacian instabilities 
are absent for ${\cal F}>0$ and ${\cal H}>0$. 
On using Eqs.~(\ref{phid}) and (\ref{dfh}), we obtain
\ba
{\cal G} &=& \frac{2\sqrt{h}\,r^4-4r^2 (\sqrt{h}-1) 
(1-h+2\sqrt{h}) \alpha-4 (\sqrt{h}+1)  (\sqrt{h}-1)^4 
\alpha^2}{r^2 \sqrt{h}\,[r^2+2(1-h)\alpha]}\,,\label{calGge} \\
{\cal F} &=& \frac{2[r^4+2r^2 (h-1) \alpha
-2(h-1)^2 \alpha^2]}{r^2[r^2+2(1-h)\alpha]}\,,\\
{\cal H} &=& \frac{2r^2+4(1-h)\alpha}{r^2}\,.\label{calHge} 
\ea
At spatial infinity we have $f=h=1-2M/r+\cdots$ and hence
all of ${\cal G}$, ${\cal F}$, and ${\cal H}$ approach 2. 

In the region close to the outer horizon $r_+=M+\sqrt{M^2-\alpha}$, 
the metric components and field kinetic term expanded around 
$r=r_+$ are given by  
\ba
\hspace{-0.5cm}
& & 
f=h=\frac{r_{+}^2-\alpha}{r_{+} (r_{+}^2+2\alpha)}(r-r_{+})
-\frac{r_{+}^6-3\alpha r_{+}^4 -6\alpha^2 r_{+}^2 -\alpha^3}
{r_{+}^2 (r_{+}^2+2\alpha)^3}(r-r_{+})^2 \nonumber \\
\hspace{-0.5cm}
& &
\qquad \qquad 
+\frac{(r_{+}^2+\alpha)(r_{+}^8 - 10\alpha r_{+}^6 - 12\alpha^2 r_{+}^4 
- 4\alpha^3 r_{+}^2 - 2\alpha^4)}
{r_{+}^3 (r_{+}^2+2\alpha)^5}(r-r_{+})^3+{\cal O}\left( (r-r_{+})^4 \right)\,,
\label{fh}\\
\hspace{-0.5cm}
& & 
X=-\frac{1}{2r_{+}^2}+\left[ \frac{r_{+}^2-\alpha}{r_{+}^5 (r_{+}^2+2\alpha)}
\right]^{1/2}(r-r_+)^{1/2}
+\frac{r_{+}^2+5\alpha}{2r_{+}^3(r_{+}^2+2\alpha)}(r-r_{+})
+{\cal O}\left( (r-r_+)^{3/2} \right)\,,
\label{Xexpan}
\ea
which are valid for $r>r_{+}$.
In the parameter range (\ref{alrange}) we have $r_{+}^2-\alpha>0$ and 
$r_{+}^2+2\alpha>0$, so the coefficient in front of $(r-r_+)^{1/2}$ in 
Eq.~(\ref{Xexpan}) is a positive real value. 
Although $X$ is finite on the horizon, it is not an analytic function 
of $r$. In Refs.~\cite{Minamitsuji:2022mlv,Minamitsuji:2022vbi} 
the analytic property of $X$ was assumed to study the BH stability 
around the horizon, so we need to handle the present case separately.
On using Eqs.~(\ref{fh})-(\ref{Xexpan}), the quantities (\ref{calGge})-(\ref{calHge}), 
which are expanded around $r=r_{+}$, reduce to
\ba
{\cal G} &=& 4\alpha \left[ \frac{r_{+}^2-\alpha}
{r_{+}^3 (r_{+}^2+2\alpha)} \right]^{1/2} \left( r-r_{+} \right)^{-1/2}
+{\cal O} \left( (r-r_{+})^0 \right)\,,
\label{cGdef2}\\
{\cal F} &=& \frac{2(r_{+}^4-2r_{+}^2 \alpha-2\alpha^2)}{r_{+}^2 (r_{+}^2+2\alpha)}
+{\cal O}\left(r-r_{+} \right)\,,
\label{cFdef2}\\
{\cal H} &=& \frac{2r_{+}^2+4\alpha}{r_{+}^2}
+{\cal O} \left( r-r_{+} \right)\,.
\label{cHdef2}
\ea
If $\alpha>0$, the ghost is absent 
in the vicinity of the outer horizon.
Under this condition we have ${\cal H}>0$, 
while the condition ${\cal F}>0$ translates to 
$r_{+}^2>(1+\sqrt{3})\alpha$, i.e., 
$\alpha<4(3\sqrt{3}-5)M^2$.
Then, the stability against odd-parity perturbations 
around $r=r_{+}$ requires that 
\be
0<\alpha<4 \left( 3\sqrt{3}-5 \right) M^2\,.
\ee
In the vicinity of $r=r_{+}$, the squared propagation speeds 
in the radial and angular directions reduce, respectively, to 
\ba
c_{r,{\rm odd}}^2
&=& \frac{2\alpha [r_+(r_+^2+2\alpha)
(r_+^2-\alpha)]^{1/2}}{r_+^4-2r_+^2 \alpha-2\alpha^2} \left( r-r_{+} \right)^{-1/2}
+{\cal O} \left( (r-r_{+})^0 \right)\,,\\
c_{\Omega,{\rm odd}}^2
&=& 2\alpha \left[ \frac{r_+(r_+^2-\alpha)}{(r_+^2+2\alpha)^3} 
\right]^{1/2}\left( r-r_{+} \right)^{-1/2}
+{\cal O} \left( (r-r_{+})^0 \right)\,.
\ea
As $r$ approaches $r_+$, both $c_{r,{\rm odd}}^2$ and 
$c_{\Omega,{\rm odd}}^2$ exhibit divergences. 
We note that the divergence of $c_{r,{\rm odd}}^2$ 
on the horizon was also reported in Ref.~\cite{Langlois:2022eta}.

In the even-parity sector, there are two dynamical variables 
$\psi$ and $\delta \phi$ corresponding to the gravitational and 
scalar-field perturbations, 
respectively \cite{Kobayashi:2014wsa,Kase:2021mix}.
Under the condition ${\cal F}>0$, the ghost is absent for
\be
{\cal K} \equiv 2{\cal P}_1-{\cal F}>0\,,
\label{Kcon}
\ee
where
\be
{\cal P}_1 \equiv \frac{h \mu}{2fr^2 {\cal H}^2} 
\left( 
\frac{fr^4 {\cal H}^4}{\mu^2 h} \right)'\,,\qquad
\mu \equiv \frac{2(\phi' a_1+r\sqrt{fh}{\cal H})}{\sqrt{fh}}\,.
\label{defP1}
\ee
The explicit form of $a_1$ is given in Appendix \ref{App}.
On using Eqs.~(\ref{phid}) and (\ref{dfh}), it follows that 
\be
{\cal K}=0\,,
\ee
at any radius.
This means that the hairy BH solution (\ref{fhal}) with the non-vanishing 
field profile (\ref{phid}) is plagued by a strong coupling problem. 
Recently, it was also found that the same strong coupling problem 
is present for NSs in 4DEGB gravity both inside and outside 
the star \cite{Minamitsuji:2022tze}.

In the limit of large frequencies, the radial propagation speed 
squared $c_{r1,{\rm even}}^2$ of gravitational perturbation 
$\psi$ is equivalent to $c_{r,{\rm odd}}^2={\cal G}/{\cal F}$. 
The condition for avoiding the Laplacian instability of  
$\delta \phi$ along the angular 
direction is \cite{Kobayashi:2014wsa,Kase:2021mix} 
\be
c_{r2,{\rm even}}^2=
\frac{2\phi'[ 4r^2 (fh)^{3/2} {\cal H} c_4 
(2\phi' a_1+r\sqrt{fh}\,{\cal H})
-2a_1^2 f^{3/2} \sqrt{h} 
\phi' {\cal G} 
+( a_1 f'+2 c_2 f ) r^2 fh 
{\cal H}^2]}{f^{5/2} h^{3/2} \mu^2 {\cal K}}>0\,,
\label{cr2even}
\ee
where $c_2$ and $c_4$ are given in Appendix \ref{App}.
Since the denominator on the right hand-side of Eq.~(\ref{cr2even}) 
is proportional to ${\cal K}$, there is the divergence of $c_{r2,{\rm even}}^2$.

In the limit of large multipoles $\ell \gg 1$, the squared propagation speeds 
of even-parity perturbations in the angular direction are \cite{Kase:2021mix}
\be
c_{\Omega \pm}^2=-B_1\pm\sqrt{B_1^2-B_2}\,,
\label{cosq}
\ee
where $B_1$ and $B_2$ are given in Appendix~A. The angular Laplacian 
instability can be avoided for $c_{\Omega \pm}^2>0$. 
These conditions can be satisfied if
\be
B_1^2 \geq B_2>0 \quad {\rm and} 
\quad B_1<0\,.
\label{B12con}
\ee
If $B_2<0$, the BH solutions are subject to the Laplacian instability 
since one of $c_{\Omega \pm}^2$ is at least negative.

As we see in Eq.~(\ref{B2def}) in Appendix \ref{App}, the denominator 
of $B_2$ is proportional to ${\cal K}$ and ${\cal F}$. 
While ${\cal K}=0$ everywhere, the 
products ${\cal K}B_2$ and ${\cal F}{\cal K}B_2$ can be finite. 
As in Refs.~\cite{Minamitsuji:2022mlv,Minamitsuji:2022vbi}, we compute 
the product ${\cal F}{\cal K}B_2$ in the vicinity of $r=r_+$.
Using the expanded solutions (\ref{fh})-(\ref{Xexpan}), 
it follows that 
\be
{\cal F}{\cal K}B_2=-\frac{\alpha^2[(2r_+^2+\alpha)^2+3\alpha^2]^2}
{r_+^2 (r_+^2+2\alpha)^4} \left( r-r_+ \right)^{-2}+
{\cal O} \left( (r-r_+)^{-3/2} \right)\,.
\ee
For $\alpha \neq 0$,  
the leading-order contribution to ${\cal F}{\cal K}B_2$ is negative 
around $r=r_+$.
This means that, even if the stability condition ${\cal F}>0$ is 
satisfied, we have ${\cal K}B_2<0$ and hence 
$B_2 \to -\infty$ for ${\cal K} \to +0$. 
Then, the BH solution (\ref{fhal}) with the field derivative 
(\ref{phid}) is inevitably plagued by the angular Laplacian 
instability around the outer horizon. 
The property ${\cal F}{\cal K}B_2<0$ also holds for hairy BH solutions where $X$ is 
an analytic function of $r$ with a non-vanishing constant $X_s$ 
on the horizon \cite{Minamitsuji:2022mlv,Minamitsuji:2022vbi}. 
Here, we have shown that the BH instability persists 
for the non-analytic function (\ref{Xexpan}) of $X$ 
in the vicinity of the outer horizon.

Let us also study the BH stability far away from the horizon ($r \gg M$). 
From Eqs.~(\ref{cr2even}) and (\ref{B2def}), we find that the ratio 
$c_{r2,{\rm even}}^2/B_2$ is finite.
Using the solution (\ref{fhal}) with (\ref{phid}) and expanding them 
at spatial infinity, it follows that 
\be
\frac{c_{r2,{\rm even}}^2}{B_2}=-2+{\cal O} 
\left( \frac{M}{r} \right)\,.
\ee
Since the leading-order term of $c_{r2,{\rm even}}^2/B_2$ is negative, 
either $c_{r2,{\rm even}}^2$ or $B_2$ is negative. 
Then, either of the stability conditions (\ref{cr2even}) or 
(\ref{B12con}) is violated far away from the horizon.
This shows that the BH solution (\ref{fhal}) is plagued by the 
instability problem not only in the vicinity of the horizon but 
also at spatial infinity.

For the specific coupling 
\be
\alpha=r_+^2\,,
\ee
there is a single horizon located at $r_{+}=M$. 
In this case, metric components $f$ and $h$ 
in Eq.~(\ref{fhal}) do not have negative values at any radius $r$. 
For this coupling, the term proportional to $r-r_+$ 
in Eq.~(\ref{fh}) and the second term on the right hand-side of 
Eq.~(\ref{Xexpan}) are vanishing.
Then, in the vicinity of $r_+=M$, we have
\be
{\cal G}=6+\frac{8\sqrt{3}}{3}+{\cal O}(r-r_+)\,,
\qquad
{\cal F}=-2+{\cal O}(r-r_+)\,,\qquad
{\cal H}=6+{\cal O}(r-r_+)\,.
\ee
Since ${\cal F}<0$ and ${\cal G}>0$ at leading order, there is the 
Laplacian instability of odd-parity perturbations 
in the radial direction. 
Moreover we have ${\cal K}=0$ and hence the strong 
coupling problem is also present in this case.

\section{Conclusions}
\label{Sec:con}

We have studied the existence and stability of BH solutions 
on a static and spherically symmetric background
in regularized 4DEGB gravity obtained from a Kaluza-Klein 
reduction of higher-dimensional EGB theory. 
The regularized 4DEGB theory derived by the rescaling 
of the GB coupling constant $\hat{\alpha} \to \alpha/(D-4)$
belongs to a subclass of Horndeski 
theories given by the action (\ref{actionre}). 
On using the background Eq.~(\ref{Jeq}) of the scalar field, 
we showed that hairy asymptotically-flat BH solutions 
respecting the boundary conditions (\ref{fhexpan}) at spatial infinity 
can exist only for a spatially-flat internal space ($\lambda=0$).

For $\lambda=0$, there is a unique hairy BH described by the 
metric components (\ref{fhal}) with the scalar derivative (\ref{phid}). 
This solution has two horizons located at $r_{\pm}=M \pm \sqrt{M^2-\alpha}$, 
where the GB coupling is in the range $-8M^2<\alpha<M^2$ for 
the existence of an outer horizon $r_+$. 
We found that the absence of ghost/Laplacian instabilities 
in the odd-parity sector requires that  $0<\alpha<4 (3\sqrt{3}-5)M^2$. 
For this hairy BH solution, the scalar field has a non-vanishing 
kinetic term $X_s$ on the horizon. This property mostly arises from  
the existence of a derivative coupling term $4 \alpha X$ 
in $G_4$. Since there is no standard kinetic term $X$ in $G_2$, 
we do not have the no-hair branch $\phi'(r)=0$ for $\alpha \neq 0$.

For the BH solution (\ref{fhal}) with (\ref{phid}), we showed that the quantity 
${\cal K}$ associated with the no-ghost condition of even-parity perturbations 
vanishes everywhere. Hence there is a strong coupling 
problem in the even-parity sector. The product ${\cal F}{\cal K}B_2$ is 
a non-vanishing finite value, but the leading-order term of ${\cal F}{\cal K}B_2$ 
is negative in the vicinity of an outer horizon. 
This means that, even if ${\cal F}$ is a positive finite value, we have 
$B_2 \to -\infty$ as ${\cal K} \to +0$. 
Then, the angular Laplacian instability of even-parity perturbations is 
present around the horizon. At spatial infinity the squared propagation 
speed $c_{r2,{\rm even}}^2$ of $\delta \phi$ divided by $B_2$ is given by 
$c_{r2,{\rm even}}^2/B_2=-2+{\cal O}(M/r)$, so 
either $c_{r2,{\rm even}}^2$ or $B_2$ is negative. 
Hence there is also the Laplacian instability of even-parity perturbations
far away from the horizon.
We note that, for $\alpha=r_+^2=M^2$, the BH has a single horizon, 
but this solution is excluded by the radial Laplacian instability of odd-parity 
perturbations in the vicinity of the horizon (i.e., ${\cal F}<0$).

Our argument of the instability of BHs is valid for the effective 4-dimensional 
action (\ref{actionre}) that belongs to a subclass of Horndeski theories. 
To discuss the stability of higher-dimensional BH solutions present 
in $D>4$ EGB theories \cite{Boulware:1985wk,Cai:2001dz}, we need 
to reformulate BH perturbations on the higher-dimensional background 
along the line of Ref.~\cite{Kodama:2003jz}.
In the presence of an electromagnetic Lagrangian $-F_{AB}F^{AB}/4$, 
it is known that there are exact charged BH solutions in $D>4$ 
dimensions \cite{Wiltshire:1985us,Wiltshire:1988uq}. 
In 4DEGB gravity reduced from $D>4$ EGB theory
with the rescaling $\hat{\alpha} \to \alpha/(D-4)$, 
we also have hairy charged BHs \cite{Fernandes:2020rpa,Hegde:2020xlv,EslamPanah:2020hoj} 
whose metric components are the same as those derived 
in the context of 4D gravity with a 
conformal anomaly \cite{Cai:2009ua,Cai:2014jea}. 
It will be of interest to investigate the stability of such charged BH solutions 
against linear perturbations, along the line of Refs.~\cite{Kase:2018voo,Kase:2018owh,Heisenberg:2018mgr} 
performed in particular classes of vector-(scalar)-tensor theories. 

\section*{Acknowledgments} 
The author thanks Masato Minamitsuji for useful comments 
on the draft. This work was supported by the Grant-in-Aid for 
Scientific Research Fund of the JSPS Nos.~19K03854 and 22K03642.

\appendix

\section{Coefficients relevant to the stability of perturbations}
\label{App}

The terms $a_1$, $c_2$, and $c_4$ appearing in Eqs.~(\ref{defP1}) 
and (\ref{cr2even}) are given by 
\ba
a_1&=&
\sqrt{fh} \left\{  \left[ G_{4,\phi}+\frac12 h ( G_{3,X}-2 G_{4,\phi X} ) \phi'^2 \right] r^2
+2 h \phi' \left[ G_{4,X}-G_{5,\phi}-\frac12h ( 2 G_{4,XX}-G_{5,\phi X} ) \phi'^2 \right] r
\right.
\notag\\
&&
\left.
+\frac12 G_{5,XX} h^3\phi'^4
-\frac12 G_{5,X} h ( 3 h-1 ) \phi'^2 
\right\}\,, \\
c_2&=&\sqrt{fh} \left[  \left\{  
\frac{1}{2f}\left( -\frac12 h ( 3 G_{3,X}-8 G_{4,\phi X} ) \phi'^2
+\frac12 h^2 ( G_{3,XX}-2 G_{4,\phi XX} ) \phi'^4
-G_{4,\phi} \right) r^2
\right.\right.
\notag\\
&&
\left.\left.
-{\frac {h\phi'}{f}} \left( 
\frac12 {h^2 ( 2 G_{4,XXX}-G_{5,\phi XX} ) \phi'^4}
-\frac12 {h ( 12 G_{4,XX}-7 G_{5,\phi X} ) \phi'^2}
+3 ( G_{4,X}-G_{5,\phi} ) \right) r
\right.\right.
\notag\\
&&
\left.\left.
+\frac{h\phi'^2}{4f}\left(
G_{5,XXX} h^3\phi'^4
- G_{5,XX} h ( 10 h-1 ) \phi'^2
+3 G_{5,X}  ( 5 h-1 ) 
\right) \right\} f'
\right.
\notag\\
&&
\left.
+\phi' \left\{ \frac12G_{2,X}-G_{3,\phi}
-\frac12 h ( G_{2,XX}-G_{3,\phi X} ) \phi'^2 \right\} r^2
-( 3 h-1 ) ( G_{4,X}-G_{5,\phi} ) \phi'
\right.
\notag\\
&&
\left.
+ 2\left\{ -\frac12h ( 3 G_{3,X}-8 G_{4,\phi X} ) \phi'^2
+\frac12h^2 ( G_{3,XX}-2 G_{4,\phi XX} ) \phi'^4
-G_{4,\phi} \right\} r
\right.
\notag\\
&&
\left.
-\frac12 h^3 ( 2 G_{4,XXX}-G_{5,\phi XX} ) \phi'^5
+\frac12 h \left\{ 2\left(6 h-1\right) G_{4,XX}+\left(1-7 h\right)G_{5,\phi X} \right\} 
\phi'^3  \right] \,,\\
c_4&=&\frac14 \frac {\sqrt {f}}{\sqrt {h}} 
\left\{ \frac {h\phi'}{f} \left[ 
2 G_{4,X}-2 G_{5,\phi}
-h ( 2 G_{4,XX}-G_{5,\phi X} ) \phi'^2
-{\frac {h\phi'  ( 3 G_{5,X}-G_{5,XX} \phi'^2h ) }{r}} \right]f'
\right.
\notag\\
&&
\left.
+4 G_{4,\phi}
+2 h ( G_{3,X}-2 G_{4,\phi X} ) \phi'^2
+{\frac {4 h ( G_{4,X}-G_{5,\phi} ) \phi'-2 h^2 ( 2 G_{4,XX}-G_{5,\phi X} ) \phi'^3}{r}} \right\}.
\ea
The quantities $B_1$ and $B_2$ are expressed as 
\ba
\hspace{-0.5cm}
&&
B_1=
\frac {r^3\sqrt {f h} {\cal H} 
(4 h \beta_0 \beta_1+\beta_2-4 \phi' a_1 \beta_3) 
-2 fh {\cal G}  [ r \sqrt{fh} {\cal K} {\cal H}  
( \beta_0+\phi' a_1 )+2\phi'^2a_1^{2}{\cal P}_1 ] }
{4f h {\cal K} {\cal H}
\beta_0^2}\,,
\label{B1def}\\
\hspace{-0.5cm}
&&
B_2=
-r^2{\frac {r^2h \beta_1 ( 2 fh {\cal F} {\cal G}\beta_0 
+r^2\beta_2) -{r}^{4}\beta_2 \beta_3
-fh{\cal F} {\cal G}  ( \phi' fh {\cal F} {\cal G}a_1 +2 r^3 \sqrt{fh} {\cal H} \beta_3 ) }
{ fh\phi' a_1 {\cal K}{\cal F} \beta_0^{2}}}\,,
\label{B2def}
\ea
where
\ba
\beta_0 &=& \phi' a_1+r\sqrt{fh} {\cal H}\,,\\
\beta_1&=&\frac12 \phi'^2 \sqrt{fh} {\cal H}e_4 
-\phi' \left(\sqrt{fh}{\cal H} \right)' c_4 
+ \frac{\sqrt{fh}}{2}\left[ \left( {\frac {f'}{f}}+{\frac {h'}{h}}-\frac{2}{r} \right) {\cal H}
+{\frac {2{\cal F}}{r}} \right] \phi' c_4+{\frac {f{\cal F} {\cal G}}{2r^2}}\,,\\
\beta_2&=& \left[ \frac{\sqrt{fh}{\cal F}}{r^2} \left( 2 hr\phi'^2c_4
+\frac{r \phi' f' \sqrt{h}}{2\sqrt{f}}{\cal H}-\phi' \sqrt{fh}\,{\cal G} \right)
-\frac{\phi' fh {\cal G}{\cal H}}{r} \left( \frac{{\cal G}'}{{\cal G}}
-\frac{{\cal H}'}{{\cal H}}+\frac{f'}{2f}-\frac{1}{r} \right) \right]a_1 
\nonumber\\
& &
-\frac{2}{r} \left( fh \right)^{3/2}{\cal F}{\cal G}{\cal H}\,, \qquad\,\,\\
\beta_3&=& \frac{\sqrt{fh}{\cal H}}{2}\phi'  
\left( hc_4'+\frac12 h' c_4-\frac{d_3}{2} \right) 
-\frac{\sqrt{fh}}{2} \left( \frac{\cal H}{r}+{\cal H}' \right) 
\left( 2 h \phi'c_4+\frac{\sqrt{fh}{\cal G}}{2r}
+\frac{f'\sqrt{h}{\cal H}}{4\sqrt{f}} \right)\notag\\
&&
+{\frac {\sqrt {fh}{\cal F}}{4r} \left(  2 h \phi'c_4
+\frac{3\sqrt{fh}{\cal G}}{r}
+\frac{f'\sqrt{h}{\cal H}}{2\sqrt{f}}
 \right) }\,,
\ea
with
\ba
d_3
&=&
-{\frac {1}{r^2} \left( {\frac {2\phi''}{\phi'}}+{\frac {h'}{h}} \right) }a_1
+{\frac {f^{3/2}h^{1/2}}{ ( f' r-2 f ) \phi'} \left( 
{\frac {2\phi''}{h\phi' r}}
+ {\frac {{f'}^{2}}{f^2}}
- {\frac {f' h'}{fh}}
-{\frac {2f'}{fr}}
+{\frac {2h'}{hr}}
+ {\frac {h'}{h^2r}} \right) }{\cal H}
\notag\\
&&
+\frac{f'r-2f}{2r} \sqrt{\frac{h}{f}} 
\frac{\partial{\cal H}}{\partial \phi}
+{\frac {\sqrt {f}}{\phi' \sqrt {h}r^2}}{\cal F}
-{\frac {{f}^{3/2}}{\sqrt {h} ( f' r-2 f ) \phi'} 
\left( {\frac {f'}{fr}}+{\frac {2\phi''}{\phi' r}}+{\frac {h'}{hr}}-\frac{2}{r^2} \right) }{\cal G}
\,,\\
e_4 &=&{\frac {1}{\phi'}}c_4'-{\frac {f'}{4f h \phi'^2}} 
\left( \sqrt{fh} {\cal H} \right)'
-{\frac {\sqrt {f}}{2\phi'^2\sqrt {h}r}}{\cal G}'
+{\frac {1}{h\phi' r^2} \left( {\frac {\phi''}{\phi'}}+\frac12 {\frac {h'}{h}} \right) }a_1
\notag\\
&&
+{\frac {\sqrt{f}}{8\sqrt{h}\phi'^2} 
\left[ {\frac { ( f' r-6 f ) f'}{f^2r}}
+\frac {h' ( f' r+4 f ) }{fhr}
-{\frac {4f ( 2 \phi'' h+h' \phi')}
{\phi' h^2r ( f' r-2 f ) }} \right] }{\cal H}
+{\frac {h'}{2h\phi'}}c_4
\notag\\
&&
-\frac{f'r-2f}{4\sqrt{fh}r\phi'}
\frac{\partial {\cal H}}
{\partial \phi}
+{\frac {f' hr-f}{2r^2\sqrt {f}{h}^{3/2}\phi'^2}}{\cal F}
+{\frac {\sqrt {f}}{2r\phi'^2{h}^{3/2}} 
\left[ {\frac {f ( 2 \phi'' h+h' \phi' ) }{h\phi'  ( f' r-2 f ) }}
+{\frac {2 f-f' hr}{2fr}} \right] }{\cal G}\,. \label{e4}
\ea

\bibliographystyle{mybibstyle}
\bibliography{bib}

\begin{thebibliography}{84}%
\makeatletter
\providecommand \@ifxundefined [1]{%
 \@ifx{#1\undefined}
}%
\providecommand \@ifnum [1]{%
 \ifnum #1\expandafter \@firstoftwo
 \else \expandafter \@secondoftwo
 \fi
}%
\providecommand \@ifx [1]{%
 \ifx #1\expandafter \@firstoftwo
 \else \expandafter \@secondoftwo
 \fi
}%
\providecommand \natexlab [1]{#1}%
\providecommand \enquote  [1]{``#1''}%
\providecommand \bibnamefont  [1]{#1}%
\providecommand \bibfnamefont [1]{#1}%
\providecommand \citenamefont [1]{#1}%
\providecommand \href@noop [0]{\@secondoftwo}%
\providecommand \href [0]{\begingroup \@sanitize@url \@href}%
\providecommand \@href[1]{\@@startlink{#1}\@@href}%
\providecommand \@@href[1]{\endgroup#1\@@endlink}%
\providecommand \@sanitize@url [0]{\catcode `\\12\catcode `\$12\catcode
  `\&12\catcode `\#12\catcode `\^12\catcode `\_12\catcode `\%12\relax}%
\providecommand \@@startlink[1]{}%
\providecommand \@@endlink[0]{}%
\providecommand \url  [0]{\begingroup\@sanitize@url \@url }%
\providecommand \@url [1]{\endgroup\@href {#1}{\urlprefix }}%
\providecommand \urlprefix  [0]{URL }%
\providecommand \Eprint [0]{\href }%
\providecommand \doibase [0]{http://dx.doi.org/}%
\providecommand \selectlanguage [0]{\@gobble}%
\providecommand \bibinfo  [0]{\@secondoftwo}%
\providecommand \bibfield  [0]{\@secondoftwo}%
\providecommand \translation [1]{[#1]}%
\providecommand \BibitemOpen [0]{}%
\providecommand \bibitemStop [0]{}%
\providecommand \bibitemNoStop [0]{.\EOS\space}%
\providecommand \EOS [0]{\spacefactor3000\relax}%
\providecommand \BibitemShut  [1]{\csname bibitem#1\endcsname}%
\let\auto@bib@innerbib\@empty
\bibitem [{\citenamefont {Copeland}\ \emph {et~al.}(2006)\citenamefont
  {Copeland}, \citenamefont {Sami},\ and\ \citenamefont
  {Tsujikawa}}]{Copeland:2006wr}%
  \BibitemOpen
  \bibfield  {author} {\bibinfo {author} {\bibfnamefont {E.~J.}\ \bibnamefont
  {Copeland}}, \bibinfo {author} {\bibfnamefont {M.}~\bibnamefont {Sami}},  and
  \bibinfo {author} {\bibfnamefont {S.}~\bibnamefont {Tsujikawa}},\ }\href
  {\doibase 10.1142/S021827180600942X} {\bibfield  {journal} {\bibinfo
  {journal} {\emph {Int. J. Mod. Phys. D}}\ }\textbf {\bibinfo {volume} {15}},\
  \bibinfo {pages} {1753} (\bibinfo {year} {2006})},\ \Eprint
  {http://arxiv.org/abs/hep-th/0603057} {arXiv:hep-th/0603057} \BibitemShut
  {NoStop}%
\bibitem [{\citenamefont {De~Felice}\ and\ \citenamefont
  {Tsujikawa}(2010)}]{DeFelice:2010aj}%
  \BibitemOpen
  \bibfield  {author} {\bibinfo {author} {\bibfnamefont {A.}~\bibnamefont
  {De~Felice}} and \bibinfo {author} {\bibfnamefont {S.}~\bibnamefont
  {Tsujikawa}},\ }\href {\doibase 10.12942/lrr-2010-3} {\bibfield  {journal}
  {\bibinfo  {journal} {\emph {Living Rev. Rel.}}\ }\textbf {\bibinfo {volume}
  {13}},\ \bibinfo {pages} {3} (\bibinfo {year} {2010})},\ \Eprint
  {http://arxiv.org/abs/1002.4928} {arXiv:1002.4928 [gr-qc]} \BibitemShut
  {NoStop}%
\bibitem [{\citenamefont {Clifton}\ \emph {et~al.}(2012)\citenamefont
  {Clifton}, \citenamefont {Ferreira}, \citenamefont {Padilla},\ and\
  \citenamefont {Skordis}}]{Clifton:2011jh}%
  \BibitemOpen
  \bibfield  {author} {\bibinfo {author} {\bibfnamefont {T.}~\bibnamefont
  {Clifton}}, \bibinfo {author} {\bibfnamefont {P.~G.}\ \bibnamefont
  {Ferreira}}, \bibinfo {author} {\bibfnamefont {A.}~\bibnamefont {Padilla}},
  and \bibinfo {author} {\bibfnamefont {C.}~\bibnamefont {Skordis}},\ }\href
  {\doibase 10.1016/j.physrep.2012.01.001} {\bibfield  {journal} {\bibinfo
  {journal} {\emph {Phys. Rept.}}\ }\textbf {\bibinfo {volume} {513}},\
  \bibinfo {pages} {1} (\bibinfo {year} {2012})},\ \Eprint
  {http://arxiv.org/abs/1106.2476} {arXiv:1106.2476 [astro-ph.CO]} \BibitemShut
  {NoStop}%
\bibitem [{\citenamefont {Joyce}\ \emph {et~al.}(2015)\citenamefont {Joyce},
  \citenamefont {Jain}, \citenamefont {Khoury},\ and\ \citenamefont
  {Trodden}}]{Joyce:2014kja}%
  \BibitemOpen
  \bibfield  {author} {\bibinfo {author} {\bibfnamefont {A.}~\bibnamefont
  {Joyce}}, \bibinfo {author} {\bibfnamefont {B.}~\bibnamefont {Jain}},
  \bibinfo {author} {\bibfnamefont {J.}~\bibnamefont {Khoury}},  and \bibinfo
  {author} {\bibfnamefont {M.}~\bibnamefont {Trodden}},\ }\href {\doibase
  10.1016/j.physrep.2014.12.002} {\bibfield  {journal} {\bibinfo  {journal}
  {\emph {Phys. Rept.}}\ }\textbf {\bibinfo {volume} {568}},\ \bibinfo {pages}
  {1} (\bibinfo {year} {2015})},\ \Eprint {http://arxiv.org/abs/1407.0059}
  {arXiv:1407.0059 [astro-ph.CO]} \BibitemShut {NoStop}%
\bibitem [{\citenamefont {Heisenberg}(2019)}]{Heisenberg:2018vsk}%
  \BibitemOpen
  \bibfield  {author} {\bibinfo {author} {\bibfnamefont {L.}~\bibnamefont
  {Heisenberg}},\ }\href {\doibase 10.1016/j.physrep.2018.11.006} {\bibfield
  {journal} {\bibinfo  {journal} {\emph {Phys. Rept.}}\ }\textbf {\bibinfo
  {volume} {796}},\ \bibinfo {pages} {1} (\bibinfo {year} {2019})},\ \Eprint
  {http://arxiv.org/abs/1807.01725} {arXiv:1807.01725 [gr-qc]} \BibitemShut
  {NoStop}%
\bibitem [{\citenamefont {Abbott}\ \emph {et~al.}(2016)\citenamefont {Abbott}
  \emph {et~al.}}]{LIGOScientific:2016aoc}%
  \BibitemOpen
  \bibfield  {author} {\bibinfo {author} {\bibfnamefont {B.~P.}\ \bibnamefont
  {Abbott}} \emph {et~al.} (\bibinfo {collaboration} {LIGO Scientific,
  Virgo}),\ }\href {\doibase 10.1103/PhysRevLett.116.061102} {\bibfield
  {journal} {\bibinfo  {journal} {\emph {Phys. Rev. Lett.}}\ }\textbf {\bibinfo
  {volume} {116}},\ \bibinfo {pages} {061102} (\bibinfo {year} {2016})},\
  \Eprint {http://arxiv.org/abs/1602.03837} {arXiv:1602.03837 [gr-qc]}
  \BibitemShut {NoStop}%
\bibitem [{\citenamefont {Berti}\ \emph {et~al.}(2015)\citenamefont {Berti}
  \emph {et~al.}}]{Berti:2015itd}%
  \BibitemOpen
  \bibfield  {author} {\bibinfo {author} {\bibfnamefont {E.}~\bibnamefont
  {Berti}} \emph {et~al.},\ }\href {\doibase 10.1088/0264-9381/32/24/243001}
  {\bibfield  {journal} {\bibinfo  {journal} {\emph {Class. Quant. Grav.}}\
  }\textbf {\bibinfo {volume} {32}},\ \bibinfo {pages} {243001} (\bibinfo
  {year} {2015})},\ \Eprint {http://arxiv.org/abs/1501.07274} {arXiv:1501.07274
  [gr-qc]} \BibitemShut {NoStop}%
\bibitem [{\citenamefont {Barack}\ \emph {et~al.}(2019)\citenamefont {Barack}
  \emph {et~al.}}]{Barack:2018yly}%
  \BibitemOpen
  \bibfield  {author} {\bibinfo {author} {\bibfnamefont {L.}~\bibnamefont
  {Barack}} \emph {et~al.},\ }\href {\doibase 10.1088/1361-6382/ab0587}
  {\bibfield  {journal} {\bibinfo  {journal} {\emph {Class. Quant. Grav.}}\
  }\textbf {\bibinfo {volume} {36}},\ \bibinfo {pages} {143001} (\bibinfo
  {year} {2019})},\ \Eprint {http://arxiv.org/abs/1806.05195} {arXiv:1806.05195
  [gr-qc]} \BibitemShut {NoStop}%
\bibitem [{\citenamefont {Horndeski}(1974)}]{Horndeski}%
  \BibitemOpen
  \bibfield  {author} {\bibinfo {author} {\bibfnamefont {G.~W.}\ \bibnamefont
  {Horndeski}},\ }\href {\doibase 10.1007/BF01807638} {\bibfield  {journal}
  {\bibinfo  {journal} {\emph {Int. J. Theor. Phys.}}\ }\textbf {\bibinfo
  {volume} {10}},\ \bibinfo {pages} {363} (\bibinfo {year} {1974})}\BibitemShut
  {NoStop}%
\bibitem [{\citenamefont {Deffayet}\ \emph {et~al.}(2011)\citenamefont
  {Deffayet}, \citenamefont {Gao}, \citenamefont {Steer},\ and\ \citenamefont
  {Zahariade}}]{Def11}%
  \BibitemOpen
  \bibfield  {author} {\bibinfo {author} {\bibfnamefont {C.}~\bibnamefont
  {Deffayet}}, \bibinfo {author} {\bibfnamefont {X.}~\bibnamefont {Gao}},
  \bibinfo {author} {\bibfnamefont {D.~A.}\ \bibnamefont {Steer}},  and
  \bibinfo {author} {\bibfnamefont {G.}~\bibnamefont {Zahariade}},\ }\href
  {\doibase 10.1103/PhysRevD.84.064039} {\bibfield  {journal} {\bibinfo
  {journal} {\emph {Phys. Rev. D}}\ }\textbf {\bibinfo {volume} {84}},\
  \bibinfo {pages} {064039} (\bibinfo {year} {2011})},\ \Eprint
  {http://arxiv.org/abs/1103.3260} {arXiv:1103.3260 [hep-th]} \BibitemShut
  {NoStop}%
\bibitem [{\citenamefont {Kobayashi}\ \emph {et~al.}(2011)\citenamefont
  {Kobayashi}, \citenamefont {Yamaguchi},\ and\ \citenamefont
  {Yokoyama}}]{KYY}%
  \BibitemOpen
  \bibfield  {author} {\bibinfo {author} {\bibfnamefont {T.}~\bibnamefont
  {Kobayashi}}, \bibinfo {author} {\bibfnamefont {M.}~\bibnamefont
  {Yamaguchi}},  and \bibinfo {author} {\bibfnamefont {J.}~\bibnamefont
  {Yokoyama}},\ }\href {\doibase 10.1143/PTP.126.511} {\bibfield  {journal}
  {\bibinfo  {journal} {\emph {Prog. Theor. Phys.}}\ }\textbf {\bibinfo
  {volume} {126}},\ \bibinfo {pages} {511} (\bibinfo {year} {2011})},\ \Eprint
  {http://arxiv.org/abs/1105.5723} {arXiv:1105.5723 [hep-th]} \BibitemShut
  {NoStop}%
\bibitem [{\citenamefont {Charmousis}\ \emph
  {et~al.}(2012{\natexlab{a}})\citenamefont {Charmousis}, \citenamefont
  {Copeland}, \citenamefont {Padilla},\ and\ \citenamefont
  {Saffin}}]{Charmousis:2011bf}%
  \BibitemOpen
  \bibfield  {author} {\bibinfo {author} {\bibfnamefont {C.}~\bibnamefont
  {Charmousis}}, \bibinfo {author} {\bibfnamefont {E.~J.}\ \bibnamefont
  {Copeland}}, \bibinfo {author} {\bibfnamefont {A.}~\bibnamefont {Padilla}},
  and \bibinfo {author} {\bibfnamefont {P.~M.}\ \bibnamefont {Saffin}},\ }\href
  {\doibase 10.1103/PhysRevLett.108.051101} {\bibfield  {journal} {\bibinfo
  {journal} {\emph {Phys. Rev. Lett.}}\ }\textbf {\bibinfo {volume} {108}},\
  \bibinfo {pages} {051101} (\bibinfo {year} {2012}{\natexlab{a}})},\ \Eprint
  {http://arxiv.org/abs/1106.2000} {arXiv:1106.2000 [hep-th]} \BibitemShut
  {NoStop}%
\bibitem [{\citenamefont {Ostrogradsky}(1850)}]{Ostrogradsky:1850fid}%
  \BibitemOpen
  \bibfield  {author} {\bibinfo {author} {\bibfnamefont {M.}~\bibnamefont
  {Ostrogradsky}},\ }\href@noop {} {\bibfield  {journal} {\bibinfo  {journal}
  {\emph {Mem. Acad. St. Petersbourg}}\ }\textbf {\bibinfo {volume} {6}},\
  \bibinfo {pages} {385} (\bibinfo {year} {1850})}\BibitemShut {NoStop}%
\bibitem [{\citenamefont {Lovelock}(1971)}]{Lovelock:1971yv}%
  \BibitemOpen
  \bibfield  {author} {\bibinfo {author} {\bibfnamefont {D.}~\bibnamefont
  {Lovelock}},\ }\href {\doibase 10.1063/1.1665613} {\bibfield  {journal}
  {\bibinfo  {journal} {\emph {J. Math. Phys.}}\ }\textbf {\bibinfo {volume}
  {12}},\ \bibinfo {pages} {498} (\bibinfo {year} {1971})}\BibitemShut
  {NoStop}%
\bibitem [{\citenamefont {Lovelock}(1972)}]{Lovelock:1972vz}%
  \BibitemOpen
  \bibfield  {author} {\bibinfo {author} {\bibfnamefont {D.}~\bibnamefont
  {Lovelock}},\ }\href {\doibase 10.1063/1.1666069} {\bibfield  {journal}
  {\bibinfo  {journal} {\emph {J. Math. Phys.}}\ }\textbf {\bibinfo {volume}
  {13}},\ \bibinfo {pages} {874} (\bibinfo {year} {1972})}\BibitemShut
  {NoStop}%
\bibitem [{\citenamefont {Padmanabhan}\ and\ \citenamefont
  {Kothawala}(2013)}]{Padmanabhan:2013xyr}%
  \BibitemOpen
  \bibfield  {author} {\bibinfo {author} {\bibfnamefont {T.}~\bibnamefont
  {Padmanabhan}} and \bibinfo {author} {\bibfnamefont {D.}~\bibnamefont
  {Kothawala}},\ }\href {\doibase 10.1016/j.physrep.2013.05.007} {\bibfield
  {journal} {\bibinfo  {journal} {\emph {Phys. Rept.}}\ }\textbf {\bibinfo
  {volume} {531}},\ \bibinfo {pages} {115} (\bibinfo {year} {2013})},\ \Eprint
  {http://arxiv.org/abs/1302.2151} {arXiv:1302.2151 [gr-qc]} \BibitemShut
  {NoStop}%
\bibitem [{\citenamefont {Fernandes}\ \emph {et~al.}(2022)\citenamefont
  {Fernandes}, \citenamefont {Carrilho}, \citenamefont {Clifton},\ and\
  \citenamefont {Mulryne}}]{Fernandes:2022zrq}%
  \BibitemOpen
  \bibfield  {author} {\bibinfo {author} {\bibfnamefont {P.~G.~S.}\
  \bibnamefont {Fernandes}}, \bibinfo {author} {\bibfnamefont {P.}~\bibnamefont
  {Carrilho}}, \bibinfo {author} {\bibfnamefont {T.}~\bibnamefont {Clifton}},
  and \bibinfo {author} {\bibfnamefont {D.~J.}\ \bibnamefont {Mulryne}},\
  }\href {\doibase 10.1088/1361-6382/ac500a} {\bibfield  {journal} {\bibinfo
  {journal} {\emph {Class. Quant. Grav.}}\ }\textbf {\bibinfo {volume} {39}},\
  \bibinfo {pages} {063001} (\bibinfo {year} {2022})},\ \Eprint
  {http://arxiv.org/abs/2202.13908} {arXiv:2202.13908 [gr-qc]} \BibitemShut
  {NoStop}%
\bibitem [{\citenamefont {Glavan}\ and\ \citenamefont
  {Lin}(2020)}]{Glavan:2019inb}%
  \BibitemOpen
  \bibfield  {author} {\bibinfo {author} {\bibfnamefont {D.}~\bibnamefont
  {Glavan}} and \bibinfo {author} {\bibfnamefont {C.}~\bibnamefont {Lin}},\
  }\href {\doibase 10.1103/PhysRevLett.124.081301} {\bibfield  {journal}
  {\bibinfo  {journal} {\emph {Phys. Rev. Lett.}}\ }\textbf {\bibinfo {volume}
  {124}},\ \bibinfo {pages} {081301} (\bibinfo {year} {2020})},\ \Eprint
  {http://arxiv.org/abs/1905.03601} {arXiv:1905.03601 [gr-qc]} \BibitemShut
  {NoStop}%
\bibitem [{\citenamefont {Arrechea}\ \emph {et~al.}(2021)\citenamefont
  {Arrechea}, \citenamefont {Delhom},\ and\ \citenamefont
  {Jim\'enez-Cano}}]{Arrechea:2020evj}%
  \BibitemOpen
  \bibfield  {author} {\bibinfo {author} {\bibfnamefont {J.}~\bibnamefont
  {Arrechea}}, \bibinfo {author} {\bibfnamefont {A.}~\bibnamefont {Delhom}},
  and \bibinfo {author} {\bibfnamefont {A.}~\bibnamefont {Jim\'enez-Cano}},\
  }\href {\doibase 10.1088/1674-1137/abc1d4} {\bibfield  {journal} {\bibinfo
  {journal} {\emph {Chin. Phys. C}}\ }\textbf {\bibinfo {volume} {45}},\
  \bibinfo {pages} {013107} (\bibinfo {year} {2021})},\ \Eprint
  {http://arxiv.org/abs/2004.12998} {arXiv:2004.12998 [gr-qc]} \BibitemShut
  {NoStop}%
\bibitem [{\citenamefont {Arrechea}\ \emph {et~al.}(2020)\citenamefont
  {Arrechea}, \citenamefont {Delhom},\ and\ \citenamefont
  {Jim\'enez-Cano}}]{Arrechea:2020gjw}%
  \BibitemOpen
  \bibfield  {author} {\bibinfo {author} {\bibfnamefont {J.}~\bibnamefont
  {Arrechea}}, \bibinfo {author} {\bibfnamefont {A.}~\bibnamefont {Delhom}},
  and \bibinfo {author} {\bibfnamefont {A.}~\bibnamefont {Jim\'enez-Cano}},\
  }\href {\doibase 10.1103/PhysRevLett.125.149002} {\bibfield  {journal}
  {\bibinfo  {journal} {\emph {Phys. Rev. Lett.}}\ }\textbf {\bibinfo {volume}
  {125}},\ \bibinfo {pages} {149002} (\bibinfo {year} {2020})},\ \Eprint
  {http://arxiv.org/abs/2009.10715} {arXiv:2009.10715 [gr-qc]} \BibitemShut
  {NoStop}%
\bibitem [{\citenamefont {Mahapatra}(2020)}]{Mahapatra:2020rds}%
  \BibitemOpen
  \bibfield  {author} {\bibinfo {author} {\bibfnamefont {S.}~\bibnamefont
  {Mahapatra}},\ }\href {\doibase 10.1140/epjc/s10052-020-08568-6} {\bibfield
  {journal} {\bibinfo  {journal} {\emph {Eur. Phys. J. C}}\ }\textbf {\bibinfo
  {volume} {80}},\ \bibinfo {pages} {992} (\bibinfo {year} {2020})},\ \Eprint
  {http://arxiv.org/abs/2004.09214} {arXiv:2004.09214 [gr-qc]} \BibitemShut
  {NoStop}%
\bibitem [{\citenamefont {G\"urses}\ \emph {et~al.}(2020)\citenamefont
  {G\"urses}, \citenamefont {\c{S}i\c{s}man},\ and\ \citenamefont
  {Tekin}}]{Gurses:2020ofy}%
  \BibitemOpen
  \bibfield  {author} {\bibinfo {author} {\bibfnamefont {M.}~\bibnamefont
  {G\"urses}}, \bibinfo {author} {\bibfnamefont {T.~c.}\ \bibnamefont
  {\c{S}i\c{s}man}},  and \bibinfo {author} {\bibfnamefont {B.}~\bibnamefont
  {Tekin}},\ }\href {\doibase 10.1140/epjc/s10052-020-8200-7} {\bibfield
  {journal} {\bibinfo  {journal} {\emph {Eur. Phys. J. C}}\ }\textbf {\bibinfo
  {volume} {80}},\ \bibinfo {pages} {647} (\bibinfo {year} {2020})},\ \Eprint
  {http://arxiv.org/abs/2004.03390} {arXiv:2004.03390 [gr-qc]} \BibitemShut
  {NoStop}%
\bibitem [{\citenamefont {Gurses}\ \emph {et~al.}(2020)\citenamefont {Gurses},
  \citenamefont {\c{S}i\c{s}man},\ and\ \citenamefont
  {Tekin}}]{Gurses:2020rxb}%
  \BibitemOpen
  \bibfield  {author} {\bibinfo {author} {\bibfnamefont {M.}~\bibnamefont
  {Gurses}}, \bibinfo {author} {\bibfnamefont {T.~c.}\ \bibnamefont
  {\c{S}i\c{s}man}},  and \bibinfo {author} {\bibfnamefont {B.}~\bibnamefont
  {Tekin}},\ }\href {\doibase 10.1103/PhysRevLett.125.149001} {\bibfield
  {journal} {\bibinfo  {journal} {\emph {Phys. Rev. Lett.}}\ }\textbf {\bibinfo
  {volume} {125}},\ \bibinfo {pages} {149001} (\bibinfo {year} {2020})},\
  \Eprint {http://arxiv.org/abs/2009.13508} {arXiv:2009.13508 [gr-qc]}
  \BibitemShut {NoStop}%
\bibitem [{\citenamefont {Fernandes}\ \emph {et~al.}(2020)\citenamefont
  {Fernandes}, \citenamefont {Carrilho}, \citenamefont {Clifton},\ and\
  \citenamefont {Mulryne}}]{Fernandes:2020nbq}%
  \BibitemOpen
  \bibfield  {author} {\bibinfo {author} {\bibfnamefont {P.~G.~S.}\
  \bibnamefont {Fernandes}}, \bibinfo {author} {\bibfnamefont {P.}~\bibnamefont
  {Carrilho}}, \bibinfo {author} {\bibfnamefont {T.}~\bibnamefont {Clifton}},
  and \bibinfo {author} {\bibfnamefont {D.~J.}\ \bibnamefont {Mulryne}},\
  }\href {\doibase 10.1103/PhysRevD.102.024025} {\bibfield  {journal} {\bibinfo
   {journal} {\emph {Phys. Rev. D}}\ }\textbf {\bibinfo {volume} {102}},\
  \bibinfo {pages} {024025} (\bibinfo {year} {2020})},\ \Eprint
  {http://arxiv.org/abs/2004.08362} {arXiv:2004.08362 [gr-qc]} \BibitemShut
  {NoStop}%
\bibitem [{\citenamefont {Hennigar}\ \emph {et~al.}(2020)\citenamefont
  {Hennigar}, \citenamefont {Kubiz\v{n}\'ak}, \citenamefont {Mann},\ and\
  \citenamefont {Pollack}}]{Hennigar:2020lsl}%
  \BibitemOpen
  \bibfield  {author} {\bibinfo {author} {\bibfnamefont {R.~A.}\ \bibnamefont
  {Hennigar}}, \bibinfo {author} {\bibfnamefont {D.}~\bibnamefont
  {Kubiz\v{n}\'ak}}, \bibinfo {author} {\bibfnamefont {R.~B.}\ \bibnamefont
  {Mann}},  and \bibinfo {author} {\bibfnamefont {C.}~\bibnamefont {Pollack}},\
  }\href {\doibase 10.1007/JHEP07(2020)027} {\bibfield  {journal} {\bibinfo
  {journal} {\emph {JHEP}}\ }\textbf {\bibinfo {volume} {07}},\ \bibinfo
  {pages} {027} (\bibinfo {year} {2020})},\ \Eprint
  {http://arxiv.org/abs/2004.09472} {arXiv:2004.09472 [gr-qc]} \BibitemShut
  {NoStop}%
\bibitem [{\citenamefont {Lu}\ and\ \citenamefont {Pang}(2020)}]{Lu:2020iav}%
  \BibitemOpen
  \bibfield  {author} {\bibinfo {author} {\bibfnamefont {H.}~\bibnamefont {Lu}}
  and \bibinfo {author} {\bibfnamefont {Y.}~\bibnamefont {Pang}},\ }\href
  {\doibase 10.1016/j.physletb.2020.135717} {\bibfield  {journal} {\bibinfo
  {journal} {\emph {Phys. Lett. B}}\ }\textbf {\bibinfo {volume} {809}},\
  \bibinfo {pages} {135717} (\bibinfo {year} {2020})},\ \Eprint
  {http://arxiv.org/abs/2003.11552} {arXiv:2003.11552 [gr-qc]} \BibitemShut
  {NoStop}%
\bibitem [{\citenamefont {Kobayashi}(2020)}]{Kobayashi:2020wqy}%
  \BibitemOpen
  \bibfield  {author} {\bibinfo {author} {\bibfnamefont {T.}~\bibnamefont
  {Kobayashi}},\ }\href {\doibase 10.1088/1475-7516/2020/07/013} {\bibfield
  {journal} {\bibinfo  {journal} {\emph {JCAP}}\ }\textbf {\bibinfo {volume}
  {07}},\ \bibinfo {pages} {013} (\bibinfo {year} {2020})},\ \Eprint
  {http://arxiv.org/abs/2003.12771} {arXiv:2003.12771 [gr-qc]} \BibitemShut
  {NoStop}%
\bibitem [{\citenamefont {Aoki}\ \emph {et~al.}(2020)\citenamefont {Aoki},
  \citenamefont {Gorji},\ and\ \citenamefont {Mukohyama}}]{Aoki:2020lig}%
  \BibitemOpen
  \bibfield  {author} {\bibinfo {author} {\bibfnamefont {K.}~\bibnamefont
  {Aoki}}, \bibinfo {author} {\bibfnamefont {M.~A.}\ \bibnamefont {Gorji}},
  and \bibinfo {author} {\bibfnamefont {S.}~\bibnamefont {Mukohyama}},\ }\href
  {\doibase 10.1016/j.physletb.2020.135843} {\bibfield  {journal} {\bibinfo
  {journal} {\emph {Phys. Lett. B}}\ }\textbf {\bibinfo {volume} {810}},\
  \bibinfo {pages} {135843} (\bibinfo {year} {2020})},\ \Eprint
  {http://arxiv.org/abs/2005.03859} {arXiv:2005.03859 [gr-qc]} \BibitemShut
  {NoStop}%
\bibitem [{\citenamefont {Boulware}\ and\ \citenamefont
  {Deser}(1985)}]{Boulware:1985wk}%
  \BibitemOpen
  \bibfield  {author} {\bibinfo {author} {\bibfnamefont {D.~G.}\ \bibnamefont
  {Boulware}} and \bibinfo {author} {\bibfnamefont {S.}~\bibnamefont {Deser}},\
  }\href {\doibase 10.1103/PhysRevLett.55.2656} {\bibfield  {journal} {\bibinfo
   {journal} {\emph {Phys. Rev. Lett.}}\ }\textbf {\bibinfo {volume} {55}},\
  \bibinfo {pages} {2656} (\bibinfo {year} {1985})}\BibitemShut {NoStop}%
\bibitem [{\citenamefont {Wheeler}(1986)}]{Wheeler:1985qd}%
  \BibitemOpen
  \bibfield  {author} {\bibinfo {author} {\bibfnamefont {J.~T.}\ \bibnamefont
  {Wheeler}},\ }\href {\doibase 10.1016/0550-3213(86)90388-3} {\bibfield
  {journal} {\bibinfo  {journal} {\emph {Nucl. Phys. B}}\ }\textbf {\bibinfo
  {volume} {273}},\ \bibinfo {pages} {732} (\bibinfo {year}
  {1986})}\BibitemShut {NoStop}%
\bibitem [{\citenamefont {Konoplya}\ and\ \citenamefont
  {Zinhailo}(2020)}]{Konoplya:2020bxa}%
  \BibitemOpen
  \bibfield  {author} {\bibinfo {author} {\bibfnamefont {R.~A.}\ \bibnamefont
  {Konoplya}} and \bibinfo {author} {\bibfnamefont {A.~F.}\ \bibnamefont
  {Zinhailo}},\ }\href {\doibase 10.1140/epjc/s10052-020-08639-8} {\bibfield
  {journal} {\bibinfo  {journal} {\emph {Eur. Phys. J. C}}\ }\textbf {\bibinfo
  {volume} {80}},\ \bibinfo {pages} {1049} (\bibinfo {year} {2020})},\ \Eprint
  {http://arxiv.org/abs/2003.01188} {arXiv:2003.01188 [gr-qc]} \BibitemShut
  {NoStop}%
\bibitem [{\citenamefont {Guo}\ and\ \citenamefont {Li}(2020)}]{Guo:2020zmf}%
  \BibitemOpen
  \bibfield  {author} {\bibinfo {author} {\bibfnamefont {M.}~\bibnamefont
  {Guo}} and \bibinfo {author} {\bibfnamefont {P.-C.}\ \bibnamefont {Li}},\
  }\href {\doibase 10.1140/epjc/s10052-020-8164-7} {\bibfield  {journal}
  {\bibinfo  {journal} {\emph {Eur. Phys. J. C}}\ }\textbf {\bibinfo {volume}
  {80}},\ \bibinfo {pages} {588} (\bibinfo {year} {2020})},\ \Eprint
  {http://arxiv.org/abs/2003.02523} {arXiv:2003.02523 [gr-qc]} \BibitemShut
  {NoStop}%
\bibitem [{\citenamefont {Fernandes}(2020)}]{Fernandes:2020rpa}%
  \BibitemOpen
  \bibfield  {author} {\bibinfo {author} {\bibfnamefont {P.~G.~S.}\
  \bibnamefont {Fernandes}},\ }\href {\doibase 10.1016/j.physletb.2020.135468}
  {\bibfield  {journal} {\bibinfo  {journal} {\emph {Phys. Lett. B}}\ }\textbf
  {\bibinfo {volume} {805}},\ \bibinfo {pages} {135468} (\bibinfo {year}
  {2020})},\ \Eprint {http://arxiv.org/abs/2003.05491} {arXiv:2003.05491
  [gr-qc]} \BibitemShut {NoStop}%
\bibitem [{\citenamefont {Konoplya}\ and\ \citenamefont
  {Zhidenko}(2020{\natexlab{a}})}]{Konoplya:2020qqh}%
  \BibitemOpen
  \bibfield  {author} {\bibinfo {author} {\bibfnamefont {R.~A.}\ \bibnamefont
  {Konoplya}} and \bibinfo {author} {\bibfnamefont {A.}~\bibnamefont
  {Zhidenko}},\ }\href {\doibase 10.1103/PhysRevD.101.084038} {\bibfield
  {journal} {\bibinfo  {journal} {\emph {Phys. Rev. D}}\ }\textbf {\bibinfo
  {volume} {101}},\ \bibinfo {pages} {084038} (\bibinfo {year}
  {2020}{\natexlab{a}})},\ \Eprint {http://arxiv.org/abs/2003.07788}
  {arXiv:2003.07788 [gr-qc]} \BibitemShut {NoStop}%
\bibitem [{\citenamefont {Hegde}\ \emph {et~al.}(2020)\citenamefont {Hegde},
  \citenamefont {Naveena~Kumara}, \citenamefont {Rizwan}, \citenamefont {M.},\
  and\ \citenamefont {Ali}}]{Hegde:2020xlv}%
  \BibitemOpen
  \bibfield  {author} {\bibinfo {author} {\bibfnamefont {K.}~\bibnamefont
  {Hegde}}, \bibinfo {author} {\bibfnamefont {A.}~\bibnamefont
  {Naveena~Kumara}}, \bibinfo {author} {\bibfnamefont {C.~L.~A.}\ \bibnamefont
  {Rizwan}}, \bibinfo {author} {\bibfnamefont {A.~K.}\ \bibnamefont {M.}},  and
  \bibinfo {author} {\bibfnamefont {M.~S.}\ \bibnamefont {Ali}},\ }\Eprint
  {http://arxiv.org/abs/2003.08778} {arXiv:2003.08778 [gr-qc]} \BibitemShut
  {NoStop}%
\bibitem [{\citenamefont {Ghosh}\ and\ \citenamefont
  {Maharaj}(2020)}]{Ghosh:2020vpc}%
  \BibitemOpen
  \bibfield  {author} {\bibinfo {author} {\bibfnamefont {S.~G.}\ \bibnamefont
  {Ghosh}} and \bibinfo {author} {\bibfnamefont {S.~D.}\ \bibnamefont
  {Maharaj}},\ }\href {\doibase 10.1016/j.dark.2020.100687} {\bibfield
  {journal} {\bibinfo  {journal} {\emph {Phys. Dark Univ.}}\ }\textbf {\bibinfo
  {volume} {30}},\ \bibinfo {pages} {100687} (\bibinfo {year} {2020})},\
  \Eprint {http://arxiv.org/abs/2003.09841} {arXiv:2003.09841 [gr-qc]}
  \BibitemShut {NoStop}%
\bibitem [{\citenamefont {Ghosh}\ and\ \citenamefont
  {Kumar}(2020)}]{Ghosh:2020syx}%
  \BibitemOpen
  \bibfield  {author} {\bibinfo {author} {\bibfnamefont {S.~G.}\ \bibnamefont
  {Ghosh}} and \bibinfo {author} {\bibfnamefont {R.}~\bibnamefont {Kumar}},\
  }\href {\doibase 10.1088/1361-6382/abc134} {\bibfield  {journal} {\bibinfo
  {journal} {\emph {Class. Quant. Grav.}}\ }\textbf {\bibinfo {volume} {37}},\
  \bibinfo {pages} {245008} (\bibinfo {year} {2020})},\ \Eprint
  {http://arxiv.org/abs/2003.12291} {arXiv:2003.12291 [gr-qc]} \BibitemShut
  {NoStop}%
\bibitem [{\citenamefont {Hosseini~Mansoori}(2021)}]{HosseiniMansoori:2020yfj}%
  \BibitemOpen
  \bibfield  {author} {\bibinfo {author} {\bibfnamefont {S.~A.}\ \bibnamefont
  {Hosseini~Mansoori}},\ }\href {\doibase 10.1016/j.dark.2021.100776}
  {\bibfield  {journal} {\bibinfo  {journal} {\emph {Phys. Dark Univ.}}\
  }\textbf {\bibinfo {volume} {31}},\ \bibinfo {pages} {100776} (\bibinfo
  {year} {2021})},\ \Eprint {http://arxiv.org/abs/2003.13382} {arXiv:2003.13382
  [gr-qc]} \BibitemShut {NoStop}%
\bibitem [{\citenamefont {Wei}\ and\ \citenamefont {Liu}(2020)}]{Wei:2020poh}%
  \BibitemOpen
  \bibfield  {author} {\bibinfo {author} {\bibfnamefont {S.-W.}\ \bibnamefont
  {Wei}} and \bibinfo {author} {\bibfnamefont {Y.-X.}\ \bibnamefont {Liu}},\
  }\href {\doibase 10.1103/PhysRevD.101.104018} {\bibfield  {journal} {\bibinfo
   {journal} {\emph {Phys. Rev. D}}\ }\textbf {\bibinfo {volume} {101}},\
  \bibinfo {pages} {104018} (\bibinfo {year} {2020})},\ \Eprint
  {http://arxiv.org/abs/2003.14275} {arXiv:2003.14275 [gr-qc]} \BibitemShut
  {NoStop}%
\bibitem [{\citenamefont {Mishra}(2020)}]{Mishra:2020gce}%
  \BibitemOpen
  \bibfield  {author} {\bibinfo {author} {\bibfnamefont {A.~K.}\ \bibnamefont
  {Mishra}},\ }\href {\doibase 10.1007/s10714-020-02763-2} {\bibfield
  {journal} {\bibinfo  {journal} {\emph {Gen. Rel. Grav.}}\ }\textbf {\bibinfo
  {volume} {52}},\ \bibinfo {pages} {106} (\bibinfo {year} {2020})},\ \Eprint
  {http://arxiv.org/abs/2004.01243} {arXiv:2004.01243 [gr-qc]} \BibitemShut
  {NoStop}%
\bibitem [{\citenamefont {Singh}\ and\ \citenamefont
  {Siwach}(2020)}]{Singh:2020xju}%
  \BibitemOpen
  \bibfield  {author} {\bibinfo {author} {\bibfnamefont {D.~V.}\ \bibnamefont
  {Singh}} and \bibinfo {author} {\bibfnamefont {S.}~\bibnamefont {Siwach}},\
  }\href {\doibase 10.1016/j.physletb.2020.135658} {\bibfield  {journal}
  {\bibinfo  {journal} {\emph {Phys. Lett. B}}\ }\textbf {\bibinfo {volume}
  {808}},\ \bibinfo {pages} {135658} (\bibinfo {year} {2020})},\ \Eprint
  {http://arxiv.org/abs/2003.11754} {arXiv:2003.11754 [gr-qc]} \BibitemShut
  {NoStop}%
\bibitem [{\citenamefont {Eslam~Panah}\ \emph {et~al.}(2020)\citenamefont
  {Eslam~Panah}, \citenamefont {Jafarzade},\ and\ \citenamefont
  {Hendi}}]{EslamPanah:2020hoj}%
  \BibitemOpen
  \bibfield  {author} {\bibinfo {author} {\bibfnamefont {B.}~\bibnamefont
  {Eslam~Panah}}, \bibinfo {author} {\bibfnamefont {K.}~\bibnamefont
  {Jafarzade}},  and \bibinfo {author} {\bibfnamefont {S.~H.}\ \bibnamefont
  {Hendi}},\ }\href {\doibase 10.1016/j.nuclphysb.2020.115269} {\bibfield
  {journal} {\bibinfo  {journal} {\emph {Nucl. Phys. B}}\ }\textbf {\bibinfo
  {volume} {961}},\ \bibinfo {pages} {115269} (\bibinfo {year} {2020})},\
  \Eprint {http://arxiv.org/abs/2004.04058} {arXiv:2004.04058 [hep-th]}
  \BibitemShut {NoStop}%
\bibitem [{\citenamefont {Yang}\ \emph {et~al.}(2020)\citenamefont {Yang},
  \citenamefont {Wan}, \citenamefont {Chen}, \citenamefont {Yang},\ and\
  \citenamefont {Wang}}]{Yang:2020czk}%
  \BibitemOpen
  \bibfield  {author} {\bibinfo {author} {\bibfnamefont {S.-J.}\ \bibnamefont
  {Yang}}, \bibinfo {author} {\bibfnamefont {J.-J.}\ \bibnamefont {Wan}},
  \bibinfo {author} {\bibfnamefont {J.}~\bibnamefont {Chen}}, \bibinfo {author}
  {\bibfnamefont {J.}~\bibnamefont {Yang}},  and \bibinfo {author}
  {\bibfnamefont {Y.-Q.}\ \bibnamefont {Wang}},\ }\href {\doibase
  10.1140/epjc/s10052-020-08511-9} {\bibfield  {journal} {\bibinfo  {journal}
  {\emph {Eur. Phys. J. C}}\ }\textbf {\bibinfo {volume} {80}},\ \bibinfo
  {pages} {937} (\bibinfo {year} {2020})},\ \Eprint
  {http://arxiv.org/abs/2004.07934} {arXiv:2004.07934 [gr-qc]} \BibitemShut
  {NoStop}%
\bibitem [{\citenamefont {Zeng}\ \emph {et~al.}(2020)\citenamefont {Zeng},
  \citenamefont {Zhang},\ and\ \citenamefont {Zhang}}]{Zeng:2020dco}%
  \BibitemOpen
  \bibfield  {author} {\bibinfo {author} {\bibfnamefont {X.-X.}\ \bibnamefont
  {Zeng}}, \bibinfo {author} {\bibfnamefont {H.-Q.}\ \bibnamefont {Zhang}},
  and \bibinfo {author} {\bibfnamefont {H.}~\bibnamefont {Zhang}},\ }\href
  {\doibase 10.1140/epjc/s10052-020-08449-y} {\bibfield  {journal} {\bibinfo
  {journal} {\emph {Eur. Phys. J. C}}\ }\textbf {\bibinfo {volume} {80}},\
  \bibinfo {pages} {872} (\bibinfo {year} {2020})},\ \Eprint
  {http://arxiv.org/abs/2004.12074} {arXiv:2004.12074 [gr-qc]} \BibitemShut
  {NoStop}%
\bibitem [{\citenamefont {Dadhich}(2020)}]{Dadhich:2020ukj}%
  \BibitemOpen
  \bibfield  {author} {\bibinfo {author} {\bibfnamefont {N.}~\bibnamefont
  {Dadhich}},\ }\href {\doibase 10.1140/epjc/s10052-020-8422-8} {\bibfield
  {journal} {\bibinfo  {journal} {\emph {Eur. Phys. J. C}}\ }\textbf {\bibinfo
  {volume} {80}},\ \bibinfo {pages} {832} (\bibinfo {year} {2020})},\ \Eprint
  {http://arxiv.org/abs/2005.05757} {arXiv:2005.05757 [gr-qc]} \BibitemShut
  {NoStop}%
\bibitem [{\citenamefont {Shaymatov}\ \emph {et~al.}(2020)\citenamefont
  {Shaymatov}, \citenamefont {Vrba}, \citenamefont {Malafarina}, \citenamefont
  {Ahmedov},\ and\ \citenamefont {Stuchl\'\i{}k}}]{Shaymatov:2020yte}%
  \BibitemOpen
  \bibfield  {author} {\bibinfo {author} {\bibfnamefont {S.}~\bibnamefont
  {Shaymatov}}, \bibinfo {author} {\bibfnamefont {J.}~\bibnamefont {Vrba}},
  \bibinfo {author} {\bibfnamefont {D.}~\bibnamefont {Malafarina}}, \bibinfo
  {author} {\bibfnamefont {B.}~\bibnamefont {Ahmedov}},  and \bibinfo {author}
  {\bibfnamefont {Z.}~\bibnamefont {Stuchl\'\i{}k}},\ }\href {\doibase
  10.1016/j.dark.2020.100648} {\bibfield  {journal} {\bibinfo  {journal} {\emph
  {Phys. Dark Univ.}}\ }\textbf {\bibinfo {volume} {30}},\ \bibinfo {pages}
  {100648} (\bibinfo {year} {2020})},\ \Eprint
  {http://arxiv.org/abs/2005.12410} {arXiv:2005.12410 [gr-qc]} \BibitemShut
  {NoStop}%
\bibitem [{\citenamefont {Arag\'on}\ \emph {et~al.}(2020)\citenamefont
  {Arag\'on}, \citenamefont {B\'ecar}, \citenamefont {Gonz\'alez},\ and\
  \citenamefont {V\'asquez}}]{Aragon:2020qdc}%
  \BibitemOpen
  \bibfield  {author} {\bibinfo {author} {\bibfnamefont {A.}~\bibnamefont
  {Arag\'on}}, \bibinfo {author} {\bibfnamefont {R.}~\bibnamefont {B\'ecar}},
  \bibinfo {author} {\bibfnamefont {P.~A.}\ \bibnamefont {Gonz\'alez}},  and
  \bibinfo {author} {\bibfnamefont {Y.}~\bibnamefont {V\'asquez}},\ }\href
  {\doibase 10.1140/epjc/s10052-020-8298-7} {\bibfield  {journal} {\bibinfo
  {journal} {\emph {Eur. Phys. J. C}}\ }\textbf {\bibinfo {volume} {80}},\
  \bibinfo {pages} {773} (\bibinfo {year} {2020})},\ \Eprint
  {http://arxiv.org/abs/2004.05632} {arXiv:2004.05632 [gr-qc]} \BibitemShut
  {NoStop}%
\bibitem [{\citenamefont {Atamurotov}\ \emph {et~al.}(2021)\citenamefont
  {Atamurotov}, \citenamefont {Shaymatov}, \citenamefont {Sheoran},\ and\
  \citenamefont {Siwach}}]{Atamurotov:2021imh}%
  \BibitemOpen
  \bibfield  {author} {\bibinfo {author} {\bibfnamefont {F.}~\bibnamefont
  {Atamurotov}}, \bibinfo {author} {\bibfnamefont {S.}~\bibnamefont
  {Shaymatov}}, \bibinfo {author} {\bibfnamefont {P.}~\bibnamefont {Sheoran}},
  and \bibinfo {author} {\bibfnamefont {S.}~\bibnamefont {Siwach}},\ }\href
  {\doibase 10.1088/1475-7516/2021/08/045} {\bibfield  {journal} {\bibinfo
  {journal} {\emph {JCAP}}\ }\textbf {\bibinfo {volume} {08}},\ \bibinfo
  {pages} {045} (\bibinfo {year} {2021})},\ \Eprint
  {http://arxiv.org/abs/2105.02214} {arXiv:2105.02214 [gr-qc]} \BibitemShut
  {NoStop}%
\bibitem [{\citenamefont {Sengupta}(2022)}]{Sengupta:2021mpf}%
  \BibitemOpen
  \bibfield  {author} {\bibinfo {author} {\bibfnamefont {S.}~\bibnamefont
  {Sengupta}},\ }\href {\doibase 10.1088/1475-7516/2022/02/020} {\bibfield
  {journal} {\bibinfo  {journal} {\emph {JCAP}}\ }\textbf {\bibinfo {volume}
  {02}},\ \bibinfo {pages} {020} (\bibinfo {year} {2022})},\ \Eprint
  {http://arxiv.org/abs/2109.10388} {arXiv:2109.10388 [gr-qc]} \BibitemShut
  {NoStop}%
\bibitem [{\citenamefont {Jaryal}\ and\ \citenamefont
  {Chatterjee}(2022)}]{Jaryal:2022rzd}%
  \BibitemOpen
  \bibfield  {author} {\bibinfo {author} {\bibfnamefont {S.~C.}\ \bibnamefont
  {Jaryal}} and \bibinfo {author} {\bibfnamefont {A.}~\bibnamefont
  {Chatterjee}},\ }\Eprint {http://arxiv.org/abs/2204.13358} {arXiv:2204.13358
  [gr-qc]} \BibitemShut {NoStop}%
\bibitem [{\citenamefont {Bravo-Gaete}\ \emph {et~al.}(2022)\citenamefont
  {Bravo-Gaete}, \citenamefont {Guajardo},\ and\ \citenamefont
  {Oliva}}]{Bravo-Gaete:2022mnr}%
  \BibitemOpen
  \bibfield  {author} {\bibinfo {author} {\bibfnamefont {M.}~\bibnamefont
  {Bravo-Gaete}}, \bibinfo {author} {\bibfnamefont {L.}~\bibnamefont
  {Guajardo}},  and \bibinfo {author} {\bibfnamefont {J.}~\bibnamefont
  {Oliva}},\ }\Eprint {http://arxiv.org/abs/2205.09282} {arXiv:2205.09282
  [hep-th]} \BibitemShut {NoStop}%
\bibitem [{\citenamefont {Fernandes}\ \emph {et~al.}(2021)\citenamefont
  {Fernandes}, \citenamefont {Carrilho}, \citenamefont {Clifton},\ and\
  \citenamefont {Mulryne}}]{Fernandes:2021ysi}%
  \BibitemOpen
  \bibfield  {author} {\bibinfo {author} {\bibfnamefont {P.~G.~S.}\
  \bibnamefont {Fernandes}}, \bibinfo {author} {\bibfnamefont {P.}~\bibnamefont
  {Carrilho}}, \bibinfo {author} {\bibfnamefont {T.}~\bibnamefont {Clifton}},
  and \bibinfo {author} {\bibfnamefont {D.~J.}\ \bibnamefont {Mulryne}},\
  }\href {\doibase 10.1103/PhysRevD.104.044029} {\bibfield  {journal} {\bibinfo
   {journal} {\emph {Phys. Rev. D}}\ }\textbf {\bibinfo {volume} {104}},\
  \bibinfo {pages} {044029} (\bibinfo {year} {2021})},\ \Eprint
  {http://arxiv.org/abs/2107.00046} {arXiv:2107.00046 [gr-qc]} \BibitemShut
  {NoStop}%
\bibitem [{\citenamefont {Regge}\ and\ \citenamefont
  {Wheeler}(1957)}]{Regge:1957td}%
  \BibitemOpen
  \bibfield  {author} {\bibinfo {author} {\bibfnamefont {T.}~\bibnamefont
  {Regge}} and \bibinfo {author} {\bibfnamefont {J.~A.}\ \bibnamefont
  {Wheeler}},\ }\href {\doibase 10.1103/PhysRev.108.1063} {\bibfield  {journal}
  {\bibinfo  {journal} {\emph {Phys. Rev.}}\ }\textbf {\bibinfo {volume}
  {108}},\ \bibinfo {pages} {1063} (\bibinfo {year} {1957})}\BibitemShut
  {NoStop}%
\bibitem [{\citenamefont {Zerilli}(1970)}]{Zerilli:1970se}%
  \BibitemOpen
  \bibfield  {author} {\bibinfo {author} {\bibfnamefont {F.~J.}\ \bibnamefont
  {Zerilli}},\ }\href {\doibase 10.1103/PhysRevLett.24.737} {\bibfield
  {journal} {\bibinfo  {journal} {\emph {Phys. Rev. Lett.}}\ }\textbf {\bibinfo
  {volume} {24}},\ \bibinfo {pages} {737} (\bibinfo {year} {1970})}\BibitemShut
  {NoStop}%
\bibitem [{\citenamefont {Kobayashi}\ \emph {et~al.}(2012)\citenamefont
  {Kobayashi}, \citenamefont {Motohashi},\ and\ \citenamefont
  {Suyama}}]{Kobayashi:2012kh}%
  \BibitemOpen
  \bibfield  {author} {\bibinfo {author} {\bibfnamefont {T.}~\bibnamefont
  {Kobayashi}}, \bibinfo {author} {\bibfnamefont {H.}~\bibnamefont
  {Motohashi}},  and \bibinfo {author} {\bibfnamefont {T.}~\bibnamefont
  {Suyama}},\ }\href {\doibase 10.1103/PhysRevD.85.084025} {\bibfield
  {journal} {\bibinfo  {journal} {\emph {Phys. Rev. D}}\ }\textbf {\bibinfo
  {volume} {85}},\ \bibinfo {pages} {084025} (\bibinfo {year} {2012})},\
  \bibinfo {note} {[Erratum:
  \href{https://doi.org/10.1103/PhysRevD.96.109903}{{\it Phys. Rev. D} {\bf
  96}, 109903 (2017)}]},\ \Eprint {http://arxiv.org/abs/1202.4893}
  {arXiv:1202.4893 [gr-qc]} \BibitemShut {NoStop}%
\bibitem [{\citenamefont {Kobayashi}\ \emph {et~al.}(2014)\citenamefont
  {Kobayashi}, \citenamefont {Motohashi},\ and\ \citenamefont
  {Suyama}}]{Kobayashi:2014wsa}%
  \BibitemOpen
  \bibfield  {author} {\bibinfo {author} {\bibfnamefont {T.}~\bibnamefont
  {Kobayashi}}, \bibinfo {author} {\bibfnamefont {H.}~\bibnamefont
  {Motohashi}},  and \bibinfo {author} {\bibfnamefont {T.}~\bibnamefont
  {Suyama}},\ }\href {\doibase 10.1103/PhysRevD.89.084042} {\bibfield
  {journal} {\bibinfo  {journal} {\emph {Phys. Rev. D}}\ }\textbf {\bibinfo
  {volume} {89}},\ \bibinfo {pages} {084042} (\bibinfo {year} {2014})},\
  \Eprint {http://arxiv.org/abs/1402.6740} {arXiv:1402.6740 [gr-qc]}
  \BibitemShut {NoStop}%
\bibitem [{\citenamefont {De~Felice}\ \emph {et~al.}(2011)\citenamefont
  {De~Felice}, \citenamefont {Suyama},\ and\ \citenamefont
  {Tanaka}}]{DeFelice:2011ka}%
  \BibitemOpen
  \bibfield  {author} {\bibinfo {author} {\bibfnamefont {A.}~\bibnamefont
  {De~Felice}}, \bibinfo {author} {\bibfnamefont {T.}~\bibnamefont {Suyama}},
  and \bibinfo {author} {\bibfnamefont {T.}~\bibnamefont {Tanaka}},\ }\href
  {\doibase 10.1103/PhysRevD.83.104035} {\bibfield  {journal} {\bibinfo
  {journal} {\emph {Phys. Rev. D}}\ }\textbf {\bibinfo {volume} {83}},\
  \bibinfo {pages} {104035} (\bibinfo {year} {2011})},\ \Eprint
  {http://arxiv.org/abs/1102.1521} {arXiv:1102.1521 [gr-qc]} \BibitemShut
  {NoStop}%
\bibitem [{\citenamefont {Motohashi}\ and\ \citenamefont
  {Suyama}(2011)}]{Motohashi:2011pw}%
  \BibitemOpen
  \bibfield  {author} {\bibinfo {author} {\bibfnamefont {H.}~\bibnamefont
  {Motohashi}} and \bibinfo {author} {\bibfnamefont {T.}~\bibnamefont
  {Suyama}},\ }\href {\doibase 10.1103/PhysRevD.84.084041} {\bibfield
  {journal} {\bibinfo  {journal} {\emph {Phys. Rev. D}}\ }\textbf {\bibinfo
  {volume} {84}},\ \bibinfo {pages} {084041} (\bibinfo {year} {2011})},\
  \Eprint {http://arxiv.org/abs/1107.3705} {arXiv:1107.3705 [gr-qc]}
  \BibitemShut {NoStop}%
\bibitem [{\citenamefont {Kase}\ \emph {et~al.}(2014)\citenamefont {Kase},
  \citenamefont {Gergely},\ and\ \citenamefont {Tsujikawa}}]{Kase:2014baa}%
  \BibitemOpen
  \bibfield  {author} {\bibinfo {author} {\bibfnamefont {R.}~\bibnamefont
  {Kase}}, \bibinfo {author} {\bibfnamefont {L.~A.}\ \bibnamefont {Gergely}},
  and \bibinfo {author} {\bibfnamefont {S.}~\bibnamefont {Tsujikawa}},\ }\href
  {\doibase 10.1103/PhysRevD.90.124019} {\bibfield  {journal} {\bibinfo
  {journal} {\emph {Phys. Rev. D}}\ }\textbf {\bibinfo {volume} {90}},\
  \bibinfo {pages} {124019} (\bibinfo {year} {2014})},\ \Eprint
  {http://arxiv.org/abs/1406.2402} {arXiv:1406.2402 [hep-th]} \BibitemShut
  {NoStop}%
\bibitem [{\citenamefont {Tattersall}\ and\ \citenamefont
  {Ferreira}(2018)}]{Tattersall:2018nve}%
  \BibitemOpen
  \bibfield  {author} {\bibinfo {author} {\bibfnamefont {O.~J.}\ \bibnamefont
  {Tattersall}} and \bibinfo {author} {\bibfnamefont {P.~G.}\ \bibnamefont
  {Ferreira}},\ }\href {\doibase 10.1103/PhysRevD.97.104047} {\bibfield
  {journal} {\bibinfo  {journal} {\emph {Phys. Rev. D}}\ }\textbf {\bibinfo
  {volume} {97}},\ \bibinfo {pages} {104047} (\bibinfo {year} {2018})},\
  \Eprint {http://arxiv.org/abs/1804.08950} {arXiv:1804.08950 [gr-qc]}
  \BibitemShut {NoStop}%
\bibitem [{\citenamefont {Kase}\ \emph {et~al.}(2020)\citenamefont {Kase},
  \citenamefont {Kimura}, \citenamefont {Sato},\ and\ \citenamefont
  {Tsujikawa}}]{Kase:2020qvz}%
  \BibitemOpen
  \bibfield  {author} {\bibinfo {author} {\bibfnamefont {R.}~\bibnamefont
  {Kase}}, \bibinfo {author} {\bibfnamefont {R.}~\bibnamefont {Kimura}},
  \bibinfo {author} {\bibfnamefont {S.}~\bibnamefont {Sato}},  and \bibinfo
  {author} {\bibfnamefont {S.}~\bibnamefont {Tsujikawa}},\ }\href {\doibase
  10.1103/PhysRevD.102.084037} {\bibfield  {journal} {\bibinfo  {journal}
  {\emph {Phys. Rev. D}}\ }\textbf {\bibinfo {volume} {102}},\ \bibinfo {pages}
  {084037} (\bibinfo {year} {2020})},\ \Eprint
  {http://arxiv.org/abs/2007.09864} {arXiv:2007.09864 [gr-qc]} \BibitemShut
  {NoStop}%
\bibitem [{\citenamefont {Khoury}\ \emph {et~al.}(2020)\citenamefont {Khoury},
  \citenamefont {Trodden},\ and\ \citenamefont {Wong}}]{Khoury:2020aya}%
  \BibitemOpen
  \bibfield  {author} {\bibinfo {author} {\bibfnamefont {J.}~\bibnamefont
  {Khoury}}, \bibinfo {author} {\bibfnamefont {M.}~\bibnamefont {Trodden}},
  and \bibinfo {author} {\bibfnamefont {S.~S.~C.}\ \bibnamefont {Wong}},\
  }\href {\doibase 10.1088/1475-7516/2020/11/044} {\bibfield  {journal}
  {\bibinfo  {journal} {\emph {JCAP}}\ }\textbf {\bibinfo {volume} {11}},\
  \bibinfo {pages} {044} (\bibinfo {year} {2020})},\ \Eprint
  {http://arxiv.org/abs/2007.01320} {arXiv:2007.01320 [astro-ph.CO]}
  \BibitemShut {NoStop}%
\bibitem [{\citenamefont {Konoplya}\ and\ \citenamefont
  {Zhidenko}(2020{\natexlab{b}})}]{Konoplya:2020juj}%
  \BibitemOpen
  \bibfield  {author} {\bibinfo {author} {\bibfnamefont {R.~A.}\ \bibnamefont
  {Konoplya}} and \bibinfo {author} {\bibfnamefont {A.}~\bibnamefont
  {Zhidenko}},\ }\href {\doibase 10.1016/j.dark.2020.100697} {\bibfield
  {journal} {\bibinfo  {journal} {\emph {Phys. Dark Univ.}}\ }\textbf {\bibinfo
  {volume} {30}},\ \bibinfo {pages} {100697} (\bibinfo {year}
  {2020}{\natexlab{b}})},\ \Eprint {http://arxiv.org/abs/2003.12492}
  {arXiv:2003.12492 [gr-qc]} \BibitemShut {NoStop}%
\bibitem [{\citenamefont {Cuyubamba}(2021)}]{Cuyubamba:2020moe}%
  \BibitemOpen
  \bibfield  {author} {\bibinfo {author} {\bibfnamefont {M.~A.}\ \bibnamefont
  {Cuyubamba}},\ }\href {\doibase 10.1016/j.dark.2021.100789} {\bibfield
  {journal} {\bibinfo  {journal} {\emph {Phys. Dark Univ.}}\ }\textbf {\bibinfo
  {volume} {31}},\ \bibinfo {pages} {100789} (\bibinfo {year} {2021})},\
  \Eprint {http://arxiv.org/abs/2004.09025} {arXiv:2004.09025 [gr-qc]}
  \BibitemShut {NoStop}%
\bibitem [{\citenamefont {Langlois}\ \emph
  {et~al.}(2022{\natexlab{a}})\citenamefont {Langlois}, \citenamefont {Noui},\
  and\ \citenamefont {Roussille}}]{Langlois:2022eta}%
  \BibitemOpen
  \bibfield  {author} {\bibinfo {author} {\bibfnamefont {D.}~\bibnamefont
  {Langlois}}, \bibinfo {author} {\bibfnamefont {K.}~\bibnamefont {Noui}},  and
  \bibinfo {author} {\bibfnamefont {H.}~\bibnamefont {Roussille}},\ }\Eprint
  {http://arxiv.org/abs/2204.04107} {arXiv:2204.04107 [gr-qc]} \BibitemShut
  {NoStop}%
\bibitem [{\citenamefont {Langlois}\ \emph
  {et~al.}(2022{\natexlab{b}})\citenamefont {Langlois}, \citenamefont {Noui},\
  and\ \citenamefont {Roussille}}]{Langlois:2022ulw}%
  \BibitemOpen
  \bibfield  {author} {\bibinfo {author} {\bibfnamefont {D.}~\bibnamefont
  {Langlois}}, \bibinfo {author} {\bibfnamefont {K.}~\bibnamefont {Noui}},  and
  \bibinfo {author} {\bibfnamefont {H.}~\bibnamefont {Roussille}},\ }\Eprint
  {http://arxiv.org/abs/2205.07746} {arXiv:2205.07746 [gr-qc]} \BibitemShut
  {NoStop}%
\bibitem [{\citenamefont {Kase}\ and\ \citenamefont
  {Tsujikawa}(2022)}]{Kase:2021mix}%
  \BibitemOpen
  \bibfield  {author} {\bibinfo {author} {\bibfnamefont {R.}~\bibnamefont
  {Kase}} and \bibinfo {author} {\bibfnamefont {S.}~\bibnamefont {Tsujikawa}},\
  }\href {\doibase 10.1103/PhysRevD.105.024059} {\bibfield  {journal} {\bibinfo
   {journal} {\emph {Phys. Rev. D}}\ }\textbf {\bibinfo {volume} {105}},\
  \bibinfo {pages} {024059} (\bibinfo {year} {2022})},\ \Eprint
  {http://arxiv.org/abs/2110.12728} {arXiv:2110.12728 [gr-qc]} \BibitemShut
  {NoStop}%
\bibitem [{\citenamefont {Minamitsuji}\ \emph
  {et~al.}(2022{\natexlab{a}})\citenamefont {Minamitsuji}, \citenamefont
  {Takahashi},\ and\ \citenamefont {Tsujikawa}}]{Minamitsuji:2022mlv}%
  \BibitemOpen
  \bibfield  {author} {\bibinfo {author} {\bibfnamefont {M.}~\bibnamefont
  {Minamitsuji}}, \bibinfo {author} {\bibfnamefont {K.}~\bibnamefont
  {Takahashi}},  and \bibinfo {author} {\bibfnamefont {S.}~\bibnamefont
  {Tsujikawa}},\ }\href {\doibase 10.1103/PhysRevD.105.104001} {\bibfield
  {journal} {\bibinfo  {journal} {\emph {Phys. Rev. D}}\ }\textbf {\bibinfo
  {volume} {105}},\ \bibinfo {pages} {104001} (\bibinfo {year}
  {2022}{\natexlab{a}})},\ \Eprint {http://arxiv.org/abs/2201.09687}
  {arXiv:2201.09687 [gr-qc]} \BibitemShut {NoStop}%
\bibitem [{\citenamefont {Minamitsuji}\ \emph
  {et~al.}(2022{\natexlab{b}})\citenamefont {Minamitsuji}, \citenamefont
  {Takahashi},\ and\ \citenamefont {Tsujikawa}}]{Minamitsuji:2022vbi}%
  \BibitemOpen
  \bibfield  {author} {\bibinfo {author} {\bibfnamefont {M.}~\bibnamefont
  {Minamitsuji}}, \bibinfo {author} {\bibfnamefont {K.}~\bibnamefont
  {Takahashi}},  and \bibinfo {author} {\bibfnamefont {S.}~\bibnamefont
  {Tsujikawa}},\ }\Eprint {http://arxiv.org/abs/2204.13837} {arXiv:2204.13837
  [gr-qc]} \BibitemShut {NoStop}%
\bibitem [{\citenamefont {Rinaldi}(2012)}]{Rinaldi:2012vy}%
  \BibitemOpen
  \bibfield  {author} {\bibinfo {author} {\bibfnamefont {M.}~\bibnamefont
  {Rinaldi}},\ }\href {\doibase 10.1103/PhysRevD.86.084048} {\bibfield
  {journal} {\bibinfo  {journal} {\emph {Phys. Rev. D}}\ }\textbf {\bibinfo
  {volume} {86}},\ \bibinfo {pages} {084048} (\bibinfo {year} {2012})},\
  \Eprint {http://arxiv.org/abs/1208.0103} {arXiv:1208.0103 [gr-qc]}
  \BibitemShut {NoStop}%
\bibitem [{\citenamefont {Anabalon}\ \emph {et~al.}(2014)\citenamefont
  {Anabalon}, \citenamefont {Cisterna},\ and\ \citenamefont
  {Oliva}}]{Anabalon:2013oea}%
  \BibitemOpen
  \bibfield  {author} {\bibinfo {author} {\bibfnamefont {A.}~\bibnamefont
  {Anabalon}}, \bibinfo {author} {\bibfnamefont {A.}~\bibnamefont {Cisterna}},
  and \bibinfo {author} {\bibfnamefont {J.}~\bibnamefont {Oliva}},\ }\href
  {\doibase 10.1103/PhysRevD.89.084050} {\bibfield  {journal} {\bibinfo
  {journal} {\emph {Phys. Rev. D}}\ }\textbf {\bibinfo {volume} {89}},\
  \bibinfo {pages} {084050} (\bibinfo {year} {2014})},\ \Eprint
  {http://arxiv.org/abs/1312.3597} {arXiv:1312.3597 [gr-qc]} \BibitemShut
  {NoStop}%
\bibitem [{\citenamefont {Minamitsuji}(2014)}]{Minamitsuji:2013ura}%
  \BibitemOpen
  \bibfield  {author} {\bibinfo {author} {\bibfnamefont {M.}~\bibnamefont
  {Minamitsuji}},\ }\href {\doibase 10.1103/PhysRevD.89.064017} {\bibfield
  {journal} {\bibinfo  {journal} {\emph {Phys. Rev. D}}\ }\textbf {\bibinfo
  {volume} {89}},\ \bibinfo {pages} {064017} (\bibinfo {year} {2014})},\
  \Eprint {http://arxiv.org/abs/1312.3759} {arXiv:1312.3759 [gr-qc]}
  \BibitemShut {NoStop}%
\bibitem [{\citenamefont {Van~Acoleyen}\ and\ \citenamefont
  {Van~Doorsselaere}(2011)}]{VanAcoleyen:2011mj}%
  \BibitemOpen
  \bibfield  {author} {\bibinfo {author} {\bibfnamefont {K.}~\bibnamefont
  {Van~Acoleyen}} and \bibinfo {author} {\bibfnamefont {J.}~\bibnamefont
  {Van~Doorsselaere}},\ }\href {\doibase 10.1103/PhysRevD.83.084025} {\bibfield
   {journal} {\bibinfo  {journal} {\emph {Phys. Rev. D}}\ }\textbf {\bibinfo
  {volume} {83}},\ \bibinfo {pages} {084025} (\bibinfo {year} {2011})},\
  \Eprint {http://arxiv.org/abs/1102.0487} {arXiv:1102.0487 [gr-qc]}
  \BibitemShut {NoStop}%
\bibitem [{\citenamefont {Charmousis}\ \emph
  {et~al.}(2012{\natexlab{b}})\citenamefont {Charmousis}, \citenamefont
  {Gouteraux},\ and\ \citenamefont {Kiritsis}}]{Charmousis:2012dw}%
  \BibitemOpen
  \bibfield  {author} {\bibinfo {author} {\bibfnamefont {C.}~\bibnamefont
  {Charmousis}}, \bibinfo {author} {\bibfnamefont {B.}~\bibnamefont
  {Gouteraux}},  and \bibinfo {author} {\bibfnamefont {E.}~\bibnamefont
  {Kiritsis}},\ }\href {\doibase 10.1007/JHEP09(2012)011} {\bibfield  {journal}
  {\bibinfo  {journal} {\emph {JHEP}}\ }\textbf {\bibinfo {volume} {09}},\
  \bibinfo {pages} {011} (\bibinfo {year} {2012}{\natexlab{b}})},\ \Eprint
  {http://arxiv.org/abs/1206.1499} {arXiv:1206.1499 [hep-th]} \BibitemShut
  {NoStop}%
\bibitem [{\citenamefont {Minamitsuji}\ and\ \citenamefont
  {Tsujikawa}(2022)}]{Minamitsuji:2022tze}%
  \BibitemOpen
  \bibfield  {author} {\bibinfo {author} {\bibfnamefont {M.}~\bibnamefont
  {Minamitsuji}} and \bibinfo {author} {\bibfnamefont {S.}~\bibnamefont
  {Tsujikawa}},\ }\Eprint {http://arxiv.org/abs/2207.04461} {arXiv:2207.04461
  [gr-qc]} \BibitemShut {NoStop}%
\bibitem [{\citenamefont {Cai}(2002)}]{Cai:2001dz}%
  \BibitemOpen
  \bibfield  {author} {\bibinfo {author} {\bibfnamefont {R.-G.}\ \bibnamefont
  {Cai}},\ }\href {\doibase 10.1103/PhysRevD.65.084014} {\bibfield  {journal}
  {\bibinfo  {journal} {\emph {Phys. Rev. D}}\ }\textbf {\bibinfo {volume}
  {65}},\ \bibinfo {pages} {084014} (\bibinfo {year} {2002})},\ \Eprint
  {http://arxiv.org/abs/hep-th/0109133} {arXiv:hep-th/0109133} \BibitemShut
  {NoStop}%
\bibitem [{\citenamefont {Kodama}\ and\ \citenamefont
  {Ishibashi}(2003)}]{Kodama:2003jz}%
  \BibitemOpen
  \bibfield  {author} {\bibinfo {author} {\bibfnamefont {H.}~\bibnamefont
  {Kodama}} and \bibinfo {author} {\bibfnamefont {A.}~\bibnamefont
  {Ishibashi}},\ }\href {\doibase 10.1143/PTP.110.701} {\bibfield  {journal}
  {\bibinfo  {journal} {\emph {Prog. Theor. Phys.}}\ }\textbf {\bibinfo
  {volume} {110}},\ \bibinfo {pages} {701} (\bibinfo {year} {2003})},\ \Eprint
  {http://arxiv.org/abs/hep-th/0305147} {arXiv:hep-th/0305147} \BibitemShut
  {NoStop}%
\bibitem [{\citenamefont {Wiltshire}(1986)}]{Wiltshire:1985us}%
  \BibitemOpen
  \bibfield  {author} {\bibinfo {author} {\bibfnamefont {D.~L.}\ \bibnamefont
  {Wiltshire}},\ }\href {\doibase 10.1016/0370-2693(86)90681-7} {\bibfield
  {journal} {\bibinfo  {journal} {\emph {Phys. Lett. B}}\ }\textbf {\bibinfo
  {volume} {169}},\ \bibinfo {pages} {36} (\bibinfo {year} {1986})}\BibitemShut
  {NoStop}%
\bibitem [{\citenamefont {Wiltshire}(1988)}]{Wiltshire:1988uq}%
  \BibitemOpen
  \bibfield  {author} {\bibinfo {author} {\bibfnamefont {D.~L.}\ \bibnamefont
  {Wiltshire}},\ }\href {\doibase 10.1103/PhysRevD.38.2445} {\bibfield
  {journal} {\bibinfo  {journal} {\emph {Phys. Rev. D}}\ }\textbf {\bibinfo
  {volume} {38}},\ \bibinfo {pages} {2445} (\bibinfo {year}
  {1988})}\BibitemShut {NoStop}%
\bibitem [{\citenamefont {Cai}\ \emph {et~al.}(2010)\citenamefont {Cai},
  \citenamefont {Cao},\ and\ \citenamefont {Ohta}}]{Cai:2009ua}%
  \BibitemOpen
  \bibfield  {author} {\bibinfo {author} {\bibfnamefont {R.-G.}\ \bibnamefont
  {Cai}}, \bibinfo {author} {\bibfnamefont {L.-M.}\ \bibnamefont {Cao}},  and
  \bibinfo {author} {\bibfnamefont {N.}~\bibnamefont {Ohta}},\ }\href {\doibase
  10.1007/JHEP04(2010)082} {\bibfield  {journal} {\bibinfo  {journal} {\emph
  {JHEP}}\ }\textbf {\bibinfo {volume} {04}},\ \bibinfo {pages} {082} (\bibinfo
  {year} {2010})},\ \Eprint {http://arxiv.org/abs/0911.4379} {arXiv:0911.4379
  [hep-th]} \BibitemShut {NoStop}%
\bibitem [{\citenamefont {Cai}(2014)}]{Cai:2014jea}%
  \BibitemOpen
  \bibfield  {author} {\bibinfo {author} {\bibfnamefont {R.-G.}\ \bibnamefont
  {Cai}},\ }\href {\doibase 10.1016/j.physletb.2014.04.044} {\bibfield
  {journal} {\bibinfo  {journal} {\emph {Phys. Lett. B}}\ }\textbf {\bibinfo
  {volume} {733}},\ \bibinfo {pages} {183} (\bibinfo {year} {2014})},\ \Eprint
  {http://arxiv.org/abs/1405.1246} {arXiv:1405.1246 [hep-th]} \BibitemShut
  {NoStop}%
\bibitem [{\citenamefont {Kase}\ \emph
  {et~al.}(2018{\natexlab{a}})\citenamefont {Kase}, \citenamefont
  {Minamitsuji}, \citenamefont {Tsujikawa},\ and\ \citenamefont
  {Zhang}}]{Kase:2018voo}%
  \BibitemOpen
  \bibfield  {author} {\bibinfo {author} {\bibfnamefont {R.}~\bibnamefont
  {Kase}}, \bibinfo {author} {\bibfnamefont {M.}~\bibnamefont {Minamitsuji}},
  \bibinfo {author} {\bibfnamefont {S.}~\bibnamefont {Tsujikawa}},  and
  \bibinfo {author} {\bibfnamefont {Y.-L.}\ \bibnamefont {Zhang}},\ }\href
  {\doibase 10.1088/1475-7516/2018/02/048} {\bibfield  {journal} {\bibinfo
  {journal} {\emph {JCAP}}\ }\textbf {\bibinfo {volume} {02}},\ \bibinfo
  {pages} {048} (\bibinfo {year} {2018}{\natexlab{a}})},\ \Eprint
  {http://arxiv.org/abs/1801.01787} {arXiv:1801.01787 [gr-qc]} \BibitemShut
  {NoStop}%
\bibitem [{\citenamefont {Kase}\ \emph
  {et~al.}(2018{\natexlab{b}})\citenamefont {Kase}, \citenamefont
  {Minamitsuji},\ and\ \citenamefont {Tsujikawa}}]{Kase:2018owh}%
  \BibitemOpen
  \bibfield  {author} {\bibinfo {author} {\bibfnamefont {R.}~\bibnamefont
  {Kase}}, \bibinfo {author} {\bibfnamefont {M.}~\bibnamefont {Minamitsuji}},
  and \bibinfo {author} {\bibfnamefont {S.}~\bibnamefont {Tsujikawa}},\ }\href
  {\doibase 10.1016/j.physletb.2018.05.078} {\bibfield  {journal} {\bibinfo
  {journal} {\emph {Phys. Lett. B}}\ }\textbf {\bibinfo {volume} {782}},\
  \bibinfo {pages} {541} (\bibinfo {year} {2018}{\natexlab{b}})},\ \Eprint
  {http://arxiv.org/abs/1803.06335} {arXiv:1803.06335 [gr-qc]} \BibitemShut
  {NoStop}%
\bibitem [{\citenamefont {Heisenberg}\ \emph {et~al.}(2018)\citenamefont
  {Heisenberg}, \citenamefont {Kase},\ and\ \citenamefont
  {Tsujikawa}}]{Heisenberg:2018mgr}%
  \BibitemOpen
  \bibfield  {author} {\bibinfo {author} {\bibfnamefont {L.}~\bibnamefont
  {Heisenberg}}, \bibinfo {author} {\bibfnamefont {R.}~\bibnamefont {Kase}},
  and \bibinfo {author} {\bibfnamefont {S.}~\bibnamefont {Tsujikawa}},\ }\href
  {\doibase 10.1103/PhysRevD.97.124043} {\bibfield  {journal} {\bibinfo
  {journal} {\emph {Phys. Rev. D}}\ }\textbf {\bibinfo {volume} {97}},\
  \bibinfo {pages} {124043} (\bibinfo {year} {2018})},\ \Eprint
  {http://arxiv.org/abs/1804.00535} {arXiv:1804.00535 [gr-qc]} \BibitemShut
  {NoStop}%
\end{thebibliography}%

\end{document}